%% file: combined.tex
\documentclass[aps,prb,onecolumn,superscriptaddress,longbibliography]{revtex4-2}

\input{preamble}

\begin{document}

\title{\papertitle}

\input{authors}

\input{abstract}

\maketitle

\makeatletter
\let\nps@addcontentsline\addcontentsline
\renewcommand{\addcontentsline}[3]{}
\makeatother

\input{main-body}

\bibliography{reference}

\makeatletter
\let\addcontentsline\nps@addcontentsline
\makeatother

\clearpage

\begin{center}
    \large\bfseries Supplementary Materials
\end{center}

\input{supplement-setup}

\input{supplement-body}

\end{document}

%% file: preamble.tex
\usepackage{amsmath}
\usepackage{amssymb}
\usepackage{graphicx}
\usepackage{bm}
\usepackage{multirow}
\usepackage{xcolor}
\usepackage{changes}
\usepackage{tikz}
\usetikzlibrary{calc, shapes.geometric}
\usepackage{algpseudocode}
\usepackage{hyperref}
\hypersetup{colorlinks=true, linkcolor=blue, citecolor=blue, urlcolor=blue}

\newcommand{\papertitle}{Extreme competition between stripes and superconductivity in the hole-doped Hubbard model}

\newcommand{\subrow}[1]{\hspace{1.2em}#1}       
\newcommand{\subsubrow}[1]{\hspace{2.4em}#1}    
\newcommand{\ampbf}[1]{\textbf{\boldmath#1}}    
\colorlet{ampgray}{black!50}
\newcommand{\ampgr}[1]{\textcolor{ampgray}{#1}} 
\newlength{\amplabwd}\newlength{\amptwd}\newlength{\ampfwd}
\newlength{\amprwd}\newlength{\ampawd}\newlength{\ampmwd}
\newlength{\amptmpwd}
\newcommand{\ampmaxw}[2]{
  \settowidth{\amptmpwd}{#2}\ifdim\amptmpwd>#1\relax\setlength{#1}{\amptmpwd}\fi}
\AtBeginDocument{%
  \setlength{\amplabwd}{0pt}\setlength{\amptwd}{0pt}\setlength{\ampfwd}{0pt}%
  \setlength{\amprwd}{0pt}\setlength{\ampawd}{0pt}\setlength{\ampmwd}{0pt}%
  \ampmaxw{\amplabwd}{\subsubrow{Final exact contraction}}%
  \ampmaxw{\amplabwd}{\ampbf{NN-fPEPS}}%
  \ampmaxw{\amptwd}{$t$ ($\mu$s)}\ampmaxw{\amptwd}{$0.00 \times 10^{0}$}%
  \ampmaxw{\amptwd}{\ampbf{$0000$}}%
  \ampmaxw{\ampfwd}{$F$ (MFLOP)}\ampmaxw{\ampfwd}{\ampbf{$0.00 \times 10^{0}$}}%
  \ampmaxw{\amprwd}{$F / t$ (TFLOP/s)}\ampmaxw{\amprwd}{\ampbf{$0.00$}}%
  \ampmaxw{\ampawd}{Arithmetic}\ampmaxw{\ampawd}{\ampbf{mixed}}\ampmaxw{\ampawd}{3$\times$TF32}%
  \ampmaxw{\ampmwd}{MFU}\ampmaxw{\ampmwd}{\ampbf{$0.0\%$}}%
  \addtolength{\amplabwd}{0.6em}\addtolength{\amptwd}{0.6em}%
  \addtolength{\ampfwd}{0.6em}\addtolength{\amprwd}{0.6em}%
  \addtolength{\ampawd}{0.6em}\addtolength{\ampmwd}{0.6em}%
}
\newcommand{\amprowstyle}[7]{%
  \makebox[\amplabwd][l]{#1{#2}} & \makebox[\amptwd][c]{#1{#3}} &
  \makebox[\ampfwd][c]{#1{#4}} & \makebox[\amprwd][c]{#1{#5}} &
  \makebox[\ampawd][c]{#1{#6}} & \makebox[\ampmwd][c]{#1{#7}}}
\newcommand{\ampplain}[1]{#1}
\newcommand{\amprow}[6]{\amprowstyle{\ampplain}{#1}{#2}{#3}{#4}{#5}{#6}}

\newcommand{\amprowgr}[6]{\amprowstyle{\ampgr}{#1}{#2}{#3}{#4}{#5}{#6}}
\newcommand{\amphead}{\amprow{}{$t$ ($\mu$s)}{$F$ (MFLOP)}{$F / t$ (TFLOP/s)}{Arithmetic}{MFU}}
\newcommand{\ampsize}[1]{\multicolumn{6}{c}{#1 lattice}}

\colorlet{tnsite}{blue!22}   
\colorlet{tnfac}{orange!33}  
\colorlet{tnproj}{green!30}  
\tikzset{
  tblob/.style={draw, fill=tnsite, rounded corners=2.5pt, minimum width=19pt, minimum height=34pt, inner sep=1pt},
  tbox/.style={draw, fill=tnfac, rounded corners=2pt, minimum width=16pt, minimum height=20pt, inner sep=1pt},
  tdia/.style={draw, fill=tnfac, diamond, minimum size=8pt, inner sep=1.2pt},
}
\newcommand{\tdg}[1]{\vcenter{\hbox{\begin{tikzpicture}[line width=0.6pt, every node/.style={font=\scriptsize}]#1\end{tikzpicture}}}}

\newcounter{algorithm}
\renewcommand{\thealgorithm}{S\arabic{algorithm}}
\newenvironment{algorithm}[1][]{%
    \par\addvspace{\intextsep}\noindent
    \minipage{\linewidth}%
    \def\caption##1{%
        \refstepcounter{algorithm}%
        \hrule height 0.8pt \nobreak \vskip 3pt
        {\raggedright \textbf{Algorithm~\thealgorithm.} ##1\par}%
        \nobreak \vskip 2pt \hrule height 0.4pt \nobreak \vskip 3pt
    }%
}{%
    \vskip 3pt \hrule height 0.8pt
    \endminipage\par\addvspace{\intextsep}%
}

%% file: authors.tex
\author{Ao Chen}
\email{chenao.phys@gmail.com}
\affiliation{Division of Chemistry and Chemical Engineering, California Institute of Technology, Pasadena, California 91125, USA}
\affiliation{Marcus Center for Theoretical Chemistry, California Institute of Technology, Pasadena, California 91125, USA}

\author{Johnnie Gray}
\affiliation{Division of Chemistry and Chemical Engineering, California Institute of Technology, Pasadena, California 91125, USA}
\affiliation{Marcus Center for Theoretical Chemistry, California Institute of Technology, Pasadena, California 91125, USA}

\author{Si-Jing Du}
\affiliation{Division of Engineering and Applied Science, California Institute of Technology, Pasadena, California 91125, USA }
\affiliation{Marcus Center for Theoretical Chemistry, California Institute of Technology, Pasadena, California 91125, USA}

\author{Garnet Kin-Lic Chan}
\email{gkc1000@gmail.com}
\affiliation{Division of Chemistry and Chemical Engineering, California Institute of Technology, Pasadena, California 91125, USA}
\affiliation{Marcus Center for Theoretical Chemistry, California Institute of Technology, Pasadena, California 91125, USA}

\date{\today}

%% file: abstract.tex

\begin{abstract}
    The two-dimensional Hubbard model was originally proposed as the minimal model of uniform $d$-wave superconductivity in the cuprate high-temperature superconductors, but recent calculations with relevant parameters in a realistic doping regime have often found other ground states.
    Further, uncertainty remains over such results due to the intrinsic bias of different computational methods. 
    Here we address this uncertainty through state-of-the-art neural tensor network and neural backflow state simulations, enabled by large-scale GPU computation, that together reach an unprecedented variational accuracy on lattices with up to 1024 sites.
    At $1/8$ doping in the pure Hubbard model, as well as with next-nearest-neighbor hopping relevant to the hole-doped cuprates, we reveal a near degeneracy among a low-energy uniform $d$-wave state, superconducting stripes, and non-superconducting stripe states on the milli-$t$ energy scale.
    Our work rebalances the narrative that the Hubbard model supports a single dominant order in relevant parameter regimes towards a picture of extreme competition between stripes and superconductivity.
\end{abstract}

%% file: main-body.tex
\section{Introduction}
The Hubbard model is a paradigmatic model of strongly correlated electrons~\cite{arovas_ARCMP_2022, Qin_ARCMP22_Hubbard}.
Despite its simple form, the two-dimensional (2D) Hubbard model is believed to capture the essential physics of high-temperature superconductivity (SC)~\cite{bednorz1986possible, Lee_RMP06_Doping-Mott}, and can be realized directly by ultracold fermionic atoms in optical lattices~\cite{Esslinger_ARCMP10_Fermi-Hubbard, Tarruell_CRP18_QS-Hubbard, Mazurenko_Nature17_AFM, Hartke_Science23_Pairing}.
In the cuprate materials that the Hubbard model aims to describe, uniform $d$-wave superconductivity is found at optimal doping, while in certain cuprates other low-temperature orders can also be observed~\cite{Keimer_Nature15_cuprates, Fradkin_RMP15_intertwined}.
For example, there is evidence for various unidirectional orders such as spin-density waves (SDW), charge-density waves (CDW), stripes (with intertwined spin and charge modulations), and even more exotic orders such as pair-density waves in several materials at low doping~\cite{Keimer_Nature15_cuprates, Fradkin_RMP15_intertwined}, and these have been argued to either coexist with, be intertwined with, or compete with superconductivity.

\begin{figure}[t]
    \centering
    \includegraphics[width=\textwidth]{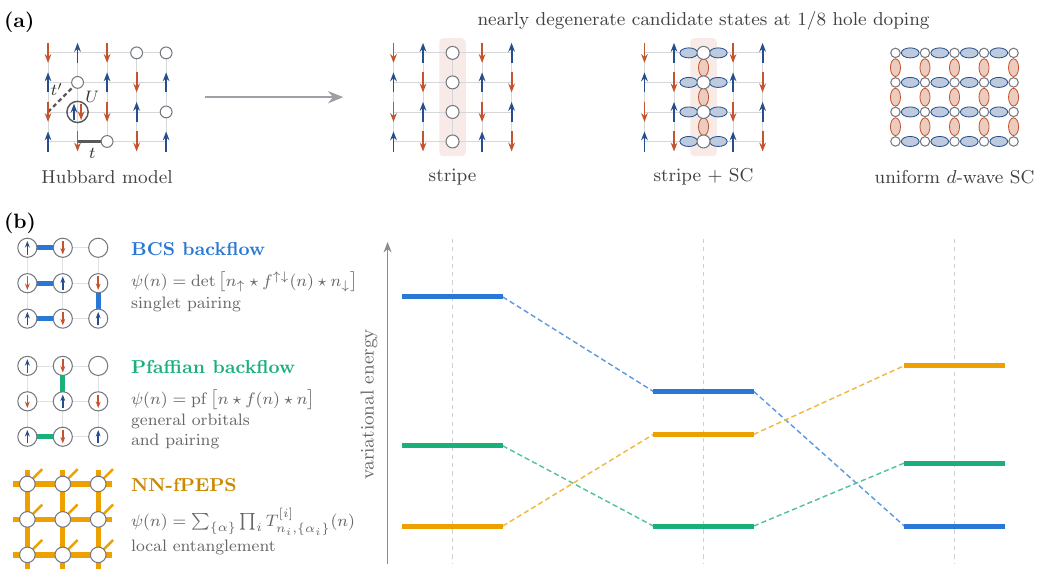}
    \caption{
        Low-energy competing states of the Hubbard model and illustrative biases of each variational wavefunction.  (a) The Hubbard model with nearest-neighbor hopping $t$, next-nearest-neighbor hopping $t'$, and on-site repulsion $U$, along with three common candidate ground states at $1/8$ hole doping.
        Recent studies report the stripe (including SDW, CDW, and their coexistence) at $t'=0$~\cite{Zheng_Science17_Hubbard, Qin_PRX20_Hubbard, Sorella_PRB23_VAFQMC, Liu_PRL25_Hubbard-fPEPS, Roth_arxiv25_HubbardSC, Jiang_npjQM26_8leg, Gu_NC26_NQS-Hubbard, Gu_PRX-Intelligence_SCALE} and $t'=-0.2$~\cite{Gu_NC26_NQS-Hubbard, Gu_PRX-Intelligence_SCALE, Ponsioen_PRB19_stripe, Jiang_npjQM26_8leg}, the stripe $+$ SC at $t'=-0.2$~\cite{Xu_Science24_Hubbard, Roth_arxiv25_HubbardSC}, and the uniform $d$-wave SC at $t'=0$~\cite{Sorella_PRB23_VAFQMC}.
        (b) The structure that carries the fermionic correlations in each ansatz (left) and, schematically, the relative variational energies of the three ansatzes for the three states in (a) (right).
    }
    \label{fig:competing_orders}
\end{figure}

Numerical studies of the hole-doped Hubbard model find different ordered ground states at the same point in the phase diagram.
We summarize some recent results here, which can be roughly categorized as shown in Fig.\,\ref{fig:competing_orders}(a).
For the pure Hubbard model, multiple methods such as the density matrix renormalization group (DMRG), projected entangled pair states (PEPS), density matrix embedding theory (DMET), constrained-path (CP-) and variational (V) auxiliary-field quantum Monte Carlo (AFQMC), and neural network backflow (NNBF) states all find a filled stripe at $U=8$ and $1/8$ doping~\cite{Zheng_Science17_Hubbard,Qin_PRX20_Hubbard, Sorella_PRB23_VAFQMC, Liu_PRL25_Hubbard-fPEPS, Roth_arxiv25_HubbardSC}, while Ref.~\cite{Sorella_PRB23_VAFQMC} also saw an energetically competitive uniform $d$-wave superconducting state at that point.
With a next-nearest-neighbor hopping $t' \approx -0.2$ relevant to the hole-doped cuprates~\cite{Andersen_JPCS95_LDA, Pavarini_PRL01_tp}, the conclusions are more varied.
Again at $1/8$ doping, robust superconductivity coexisting with partially filled stripes has been reported in a DMRG/CP-AFQMC study~\cite{Xu_Science24_Hubbard} and a neural Pfaffian study~\cite{Roth_arxiv25_HubbardSC}, whereas other simulations using infinite PEPS~\cite{Ponsioen_PRB19_stripe} and other NNBF states~\cite{Gu_NC26_NQS-Hubbard, Gu_PRX-Intelligence_SCALE} have found non-superconducting stripes, while a DMRG study found only a CDW with short-range pairing~\cite{Jiang_npjQM26_8leg}.

The diversity of observations can be traced to the sensitivity of the Hubbard model ground state to biases of the numerical simulations. While it is possible to make progress through the consensus of different techniques which individually have uncertainties larger than the energy scale of competing orders---for example, the study of stripes in Ref.~\cite{Zheng_Science17_Hubbard} involved an energy spread across methods of roughly $0.01$~$t$, larger than the energy difference between competing stripe wavelengths---the only way to eliminate this uncertainty is to improve the absolute accuracy of the simulations. To this end, we employ variational wavefunctions that can in principle be improved to arbitrary accuracy and that in practice we push to the precision of a few milli-$t$ on lattices with up to 1024 sites.
We employ two variants of NNBF~\cite{Luo_PRL19_Backflow}, which have recently demonstrated high accuracy in Hubbard model studies~\cite{Gu_NC26_NQS-Hubbard, Chen_PNAS26_HFPS, Roth_arxiv25_HubbardSC, Loehr_arxiv25_NNBF, Gu_PRX-Intelligence_SCALE, Rende_arxiv26_SBP}, as well as a newly introduced neural tensor network state~\cite{Du_PRB26_NN-fTNS}, namely the neural-network fermionic PEPS (NN-fPEPS). The rationale for using multiple variational forms is that each relies on an underlying fermionic ansatz that may impose a bias in the calculation. The Bardeen--Cooper--Schrieffer (BCS)~\cite{Bardeen_PR57_BCS} and Pfaffian NNBF states have been argued to favor paired states~\cite{Viteritti_arxiv26_bias}, while the fermionic PEPS (fPEPS)~\cite{Corboz_PRB10_fPEPS, Kraus_PRA10_fPEPS, Mortier_SP25_fTN,Gao_PRR25_fTN-contraction} underlying the NN-fPEPS wavefunction favors low entanglement, a very different starting point.
In each case, through intensive optimization of the underlying computational infrastructure, as well as through deploying large-scale graphics processing unit (GPU) computation, we push each variational form to the practical limits of accuracy. As we shall detail below, this leads to new findings and insights regarding the ground-state order in the hole-doped Hubbard model.

\section{Hubbard model and competing orders}
\label{sec:model}
The Hamiltonian of the Hubbard model is
\begin{equation}
    H = -t \sum_{\langle i, j \rangle, \sigma} \left( c^\dagger_{i\sigma} c_{j\sigma} + \mathrm{H.c.} \right)
    - t' \sum_{\langle\langle i, j \rangle\rangle, \sigma} \left( c^\dagger_{i\sigma} c_{j\sigma} + \mathrm{H.c.} \right)
    + U \sum_i n_{i\uparrow} n_{i\downarrow},
    \label{eq:hubbard}
\end{equation}
where $c^\dagger_{i\sigma}$ ($c_{i\sigma}$) creates (annihilates) an electron with spin $\sigma \in \{\uparrow, \downarrow\}$ on site $i$, $n_{i\sigma} = c^\dagger_{i\sigma} c_{i\sigma}$, $\langle i, j \rangle$ and $\langle\langle i, j \rangle\rangle$ denote nearest- and next-nearest-neighbor bonds with hopping amplitudes $t$ and $t'$, H.c. denotes the Hermitian conjugate, and $U$ is the on-site repulsion.
In this work, we take $t = 1$ as the unit of energy and focus on square lattices of $N = L \times L$ sites with periodic boundary conditions (PBC) at $1/8$ hole doping and $U = 8$, for which there have been many recent studies~\cite{Zheng_Science17_Hubbard, Qin_PRX20_Hubbard, Sorella_PRB23_VAFQMC, Xu_Science24_Hubbard, Roth_arxiv25_HubbardSC, Gu_NC26_NQS-Hubbard, Gu_PRX-Intelligence_SCALE, Rende_arxiv26_SBP, Viteritti_arxiv26_bias}.

We characterize the ground state by two order parameters, the $d$-wave pairing $\Delta_{\mathrm{SC}}^2$ and the stripe order $M^2_{\mathrm{str}}$, which compete in this regime.
The pairing order parameter is extracted from the long-distance behavior of the pair-pair correlation function
\begin{equation}
    \begin{split}
        C_p(r) &= \langle \Delta^\dagger_0 \Delta_r \rangle - \mathcal{N}_{0,r} \,, \\
        \Delta^\dagger_r &= \frac{1}{4} \sum_{\eta} \frac{h_{\eta}}{\sqrt{2}} \left( c^\dagger_{r \uparrow} c^\dagger_{r + \eta, \downarrow} - c^\dagger_{r \downarrow} c^\dagger_{r + \eta, \uparrow} \right) \,,
    \end{split}
    \label{eq:pairing}
\end{equation}
where $r$ labels lattice sites, $\eta$ runs over the four nearest-neighbor vectors, the form factor $h_\eta = +1$ ($-1$) on horizontal (vertical) bonds selects the $d_{x^2 - y^2}$ channel in the convention of Refs.~\cite{Qin_PRX20_Hubbard, Xu_Science24_Hubbard, Rende_arxiv26_SBP}, and $\mathcal{N}_{0,r}$ removes the disconnected normal-state contribution, namely the products of one-body correlators $\langle c^\dagger c \rangle \langle c^\dagger c \rangle$ in the Wick decomposition of $\langle \Delta^\dagger_0 \Delta_r \rangle$~\cite{Roth_arxiv25_HubbardSC}.
The $d$-wave order parameter is
\begin{equation}
    \Delta_{\mathrm{SC}}^2 = \frac{1}{\mathcal{M}} \sum_{|r| \geq r_{\max}} C_p(r),
    \label{eq:delta_SC}
\end{equation}
which averages $C_p(r)$ over the $\mathcal{M}$ vectors with $|r| \geq r_{\max}$, where $r_{\max}$ is half the largest distance on the PBC lattice~\cite{Rende_arxiv26_SBP}.
The stripe order parameter is defined through the spin structure factor $S(k) = \sum_r e^{i k \cdot r} \langle \bm{S}_0 \cdot \bm{S}_r \rangle$ as
\begin{equation}
    M^2_{\mathrm{str}} = \frac{1}{4N} \sum_{\alpha = 1}^{4} S(k_\alpha) \,,
    \label{eq:stripe}
\end{equation}
where $\bm{S}_r$ is the spin operator at site $r$ and $k_\alpha \in \{(\pi \pm q, \pi), (\pi, \pi \pm q)\}$ are the four peak wavevectors of the stripe order~\cite{Viteritti_arxiv26_bias}. Apart from the SDW represented by $M^2_{\mathrm{str}}$, the stripes also carry a CDW, which coexists with the SDW but appears to be much weaker in our calculations, and is discussed in Sec.\,VIII of the Supplementary Materials (SM)~\cite{SM}.

The two order parameters can be used to distinguish three candidate ground states, namely the stripe state with $M^2_{\mathrm{str}} > 0$ and $\Delta^2_{\mathrm{SC}} = 0$, the uniform $d$-wave superconductor with $\Delta^2_{\mathrm{SC}} > 0$ and $M^2_{\mathrm{str}} = 0$, and a striped $d$-wave superconductor where both orders are nonzero.
Both order parameters must be extrapolated to the thermodynamic limit (TDL).
Since each of them breaks a continuous symmetry, they are associated with gapless modes, such as the phase mode of the superconductor and the spin waves of the stripe, which make correlations approach their long-distance value with a correction $\propto 1/|r|$, as in the Heisenberg antiferromagnet~\cite{Sandvik_PRB97_QMCHeisenberg}.
Hence the leading finite-size corrections to $\Delta^2_{\mathrm{SC}}$ and $M^2_{\mathrm{str}}$ are linear in $1/L$, and we extrapolate both by straight lines in $1/L$.
This linear form of $M^2_{\mathrm{str}}$ against $1/L$ assumes long-range order.
In a state without stripe order, the spin correlations decay over a finite correlation length $\xi$, so that for $L \gg \xi$ the structure factor $S(k_\alpha)$ no longer grows with the system size and $M^2_{\mathrm{str}} \propto 1/N = 1/L^2$.
We therefore extrapolate the stripe order of states whose $S(k_\alpha)$ saturates with the size by $M^2_{\mathrm{str}} = a + b/L^2$.
We evaluate these orders for the three types of wavefunctions introduced in Sec.\,\ref{sec:wavefunctions}, for the pure Hubbard model at $t' = 0$ (Sec.\,\ref{sec:pure}) and for the $t$--$t'$ Hubbard model at $t' = -0.2$ (Sec.\,\ref{sec:ttp}).

\section{Neural and tensor network wavefunctions}
\label{sec:wavefunctions}
In variational Monte Carlo (VMC) simulations~\cite{McMillan_PR65_VMC, Becca_17_VMCtext} of the Hubbard model with neural quantum states (NQS)~\cite{Carleo_Science17_NQS}, the variational state is usually expanded in the occupation basis as $|\Psi\rangle = \sum_n \psi(n) |n\rangle$, where the configuration space of each site is $n_i \in \{0, \uparrow, \downarrow, \uparrow\downarrow\}$.
The NNBF wavefunctions we use are based on the symmetry-preserving backflow pairing state of Ref.~\cite{Rende_arxiv26_SBP}.
We refer to the specific pairing form in Ref.~\cite{Rende_arxiv26_SBP} as the BCS backflow state, and we further generalize it to a Pfaffian pairing form~\cite{Bouchaud_JPF88_pfaffian, bajdich2006pfaffian, bajdich2008pfaffian}, the Pfaffian backflow state, which allows for general orbital coefficients and pairing among the $N_e$ electrons.
Its amplitude is
\begin{equation}
    \psi_{\mathrm{Pf\text{-}BF}}(n) = \operatorname{pf} \big[ n \star f(n) \star n \big], \qquad
    f^{\sigma\sigma'}_{ij}(n) = p_{ij} \sum_{\alpha=1}^{d} W^{\sigma\sigma'}_{i-j,\alpha}\, y^{\sigma}_{i\alpha}(n)\, y^{\sigma'}_{j\alpha}(n),
    \label{eq:sbp_pf}
\end{equation}
where $f(n)$ is an antisymmetric $2N \times 2N$ pairing matrix whose element $f^{\sigma\sigma'}_{ij}(n)$ connects the spin-orbitals $(i, \sigma)$ and $(j, \sigma')$, and $n \star f(n) \star n$ denotes its $N_e \times N_e$ submatrix obtained by retaining only the rows and columns of occupied spin-orbitals.
The backflow features $y^{\sigma}_{i}(n)$ are $d$-dimensional vectors generated by a residual convolutional neural network (CNN) detailed in SM Sec.\,I~\cite{SM}, with $y^{\uparrow}$ and $y^{\downarrow}$ given by the two halves of its output channels, and $p_{ij} \propto e^{-\lambda d(i, j)}$ is a normalized envelope that decays with the distance $d(i, j)$ between the two sites at a trainable rate $\lambda$.
The pairing weights $W^{\sigma\sigma'}_{i-j,\alpha}$ depend only on the displacement $i - j$, so $\psi_{\mathrm{Pf\text{-}BF}}(n)$ is translation invariant by construction without any symmetry projection, and $W^{\sigma'\sigma}_{-r} = -W^{\sigma\sigma'}_{r}$ so that $f(n)$ is exactly antisymmetric for the symmetric envelope $p_{ij} = p_{ji}$.

The BCS backflow state is a special case of the Pfaffian backflow state in Eq.\,\eqref{eq:sbp_pf} with same-spin weights $W^{\uparrow\uparrow} = W^{\downarrow\downarrow} = 0$.
The pairing is then restricted to opposite spins and the Pfaffian reduces to a determinant up to a configuration-independent sign,
\begin{equation}
    \psi_{\mathrm{BCS\text{-}BF}}(n) = \det \big[ n_\uparrow \star f^{\uparrow\downarrow}(n) \star n_\downarrow \big],
    \label{eq:sbp_det}
\end{equation}
where $n_\uparrow \star f^{\uparrow\downarrow}(n) \star n_\downarrow$ retains the rows of occupied spin-up sites and the columns of occupied spin-down sites of the $N \times N$ block $f^{\uparrow\downarrow}(n)$.
The two forms share the same network architecture, and the Pfaffian backflow only adds the same-spin weights.
In our reimplementation of the BCS backflow, the Transformer backbone of Ref.~\cite{Rende_arxiv26_SBP} is replaced by the same residual CNN.

\begin{figure}[t]
    \centering
    \includegraphics[width=\textwidth]{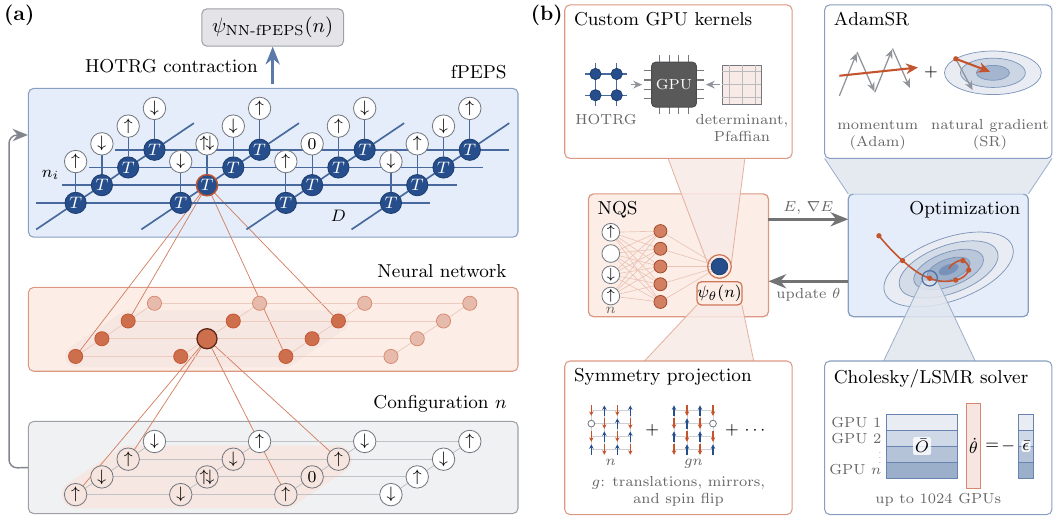}
    \caption{
        NN-fPEPS ansatz and optimization loop.
        (a) A neural network (middle, much deeper in practice) maps the configuration $n$ (bottom) to the local tensors $T^{[i]}(n)$ of the fPEPS (top), whose higher-order tensor renormalization group (HOTRG) contraction gives $\psi_{\text{NN-fPEPS}}(n)$.
        The backflow states replace the fPEPS by the Pfaffian or determinant [Eqs.\,\eqref{eq:sbp_pf} and \eqref{eq:sbp_det}].
        (b) Monte Carlo sampling of the NQS $\psi_\theta(n)$ yields $E$ and $\nabla E$, from which the optimization updates the variational parameters $\theta$.
        Zoomed-in boxes show other important computational components: custom GPU kernels, symmetry projection of the states, the adaptive moment stochastic reconfiguration (AdamSR) optimizer combining momentum with the natural gradient, and the Cholesky or least-squares minimal residual (LSMR) solver used at every iteration.
    }
    \label{fig:nn_fpeps}
\end{figure}

Our NN-fPEPS is instead built on the fPEPS~\cite{Corboz_PRB10_fPEPS, Kraus_PRA10_fPEPS, Mortier_SP25_fTN}, whose amplitude is given by
\begin{equation}
    \psi_{\mathrm{fPEPS}}(n) = \sum_{\{\alpha\}} \prod_i T^{[i]}_{n_i, \{\alpha_i\}},
    \label{eq:fpeps}
\end{equation}
where $T^{[i]}$ is the local tensor at site $i$, the subscript $n_i$ gives the physical index slice of the tensor at the local configuration, and $\{\alpha_i\}$ denotes the set of its four virtual indices of bond dimension $D$, each shared with a nearest neighbor and traced out by the sum over all virtual indices $\{\alpha\}$.
The tensors are $\mathbb{Z}_2$-graded such that the contraction automatically generates the minus signs required by fermionic anticommutation~\cite{Corboz_PRB10_fPEPS, Kraus_PRA10_fPEPS, Mortier_SP25_fTN,Gao_PRR25_fTN-contraction}.
Variational fPEPS, optimized either by simple update or by gradient optimization, have previously been shown to deliver good accuracy at large $D$ in simulations of the Hubbard model~\cite{Liu_PRL25_Hubbard-fPEPS}.
Below we will use variational fPEPS, with PBC and bond dimension extrapolation, to give an independent estimate of the exact ground-state energy on the $16 \times 16$ lattice at $t'=0$ (see SM Sec.\,VI~\cite{SM} for details on the pure fPEPS PBC contraction and energy evaluation strategy).

Building on Ref.~\cite{Du_PRB26_NN-fTNS}, which introduced a neural parametrization (neuralization) of tensor networks~\cite{chen2023antn,Du_PRB26_NN-fTNS,Fan_arxiv26_neuralTN} for fermions,
we promote the fPEPS to an NN-fPEPS by generating all local tensors with a neural network, and the amplitude becomes
\begin{equation}
    \psi_{\text{NN-fPEPS}}(n) = \sum_{\{\alpha\}} \prod_i T^{[i]}_{n_i, \{\alpha_i\}}(n).
    \label{eq:neural-peps}
\end{equation}
As in the backflow form, the configuration $n$ enters the amplitude in two distinct ways. The argument $(n)$ indicates that the tensor elements are generated by the neural network from the global configuration, while the subscript $n_i$ slices the physical index of each tensor in the contraction, as in Eq.\,\eqref{eq:fpeps} and in analogy with the selection of occupied spin-orbitals by $n \star f(n) \star n$ in Eq.\,\eqref{eq:sbp_pf}.
The NN-fPEPS architecture is illustrated in Fig.\,\ref{fig:nn_fpeps}(a), and, as in our backflow states, the tensors are generated by a residual CNN (SM Sec.\,I~\cite{SM}).

The tensor contraction in Eq.\,\eqref{eq:neural-peps} is performed approximately by the higher-order tensor renormalization group (HOTRG)~\cite{Xie_PRB12_HOTRG, Iino_PRB19_BoundaryTRG}, which repeatedly merges pairs of adjacent rows, compresses the doubled bonds back to dimension $\chi$ by applying oblique projectors, and is naturally compatible with PBC (SM Sec.\,II~\cite{SM}).
By keeping the same (set of) contraction path(s) for every configuration $n$, the approximate HOTRG contraction ensures strict variationality~\cite{Liu_PRL24_PEPS-VMC} for any $D, \chi$, and can be easily parallelized~\cite{Du_arxiv26_TN-VMC}.
Its time complexity per amplitude is $\mathcal{O}(N \chi^7)$, or $\mathcal{O}(N)$ at fixed $D$ and $\chi$.
To avoid the high cost of large $D$ and $\chi$, we choose $D = \chi = 4$ throughout this work and rely on the neural network rather than the tensor network to increase the expressivity.
To further restore the strict translation symmetry~\cite{Rende_arxiv26_SBP} broken by HOTRG, we express $\psi_{\text{NN-fPEPS}}(n)$ as a superposition of a set of contraction paths, which leads to a slightly higher computational complexity of $\mathcal{O}(N \log N)$ (SM Sec.\,III~\cite{SM}). The NN-fPEPS thus avoids the $\mathcal{O}(N^3)$ cost of the Pfaffian and the determinant in the backflow calculations at large system sizes.

The NN-fPEPS and the backflow wavefunctions retain different inductive biases arising from how the fermionic sign structure is enforced, as illustrated in Fig.\,\ref{fig:competing_orders}(b).
The BCS and Pfaffian backflow states capture pairing through the mean-field wavefunction and have been shown to reach high accuracy if the ground state is indeed a BCS superconductor~\cite{Chen_PNAS26_HFPS,Rende_arxiv26_SBP}.
Empirically, however, it has been argued that these reference forms bias the state toward superconductivity~\cite{Viteritti_arxiv26_bias}.
In addition, Sec.\,\ref{sec:results} includes the determinant backflow state (ACE) of Ref.~\cite{Gu_PRX-Intelligence_SCALE}, which potentially favors stripes over superconductivity~\cite{Viteritti_arxiv26_bias}.
The fPEPS in the NN-fPEPS, in contrast to backflow wavefunctions, favors locally entangled (area-law) ground states~\cite{Verstraete_AiP08_DMRG, Eisert_RMP10_AreaLaw}. A priori, this feature favors neither stripes nor superconductivity, although specific stripe or superconducting states may carry different amounts of entanglement.

Our simulations are implemented in \textsc{Quantax}~\cite{quantax}, which is built on \textsc{Jax}~\cite{jax2018github} and \textsc{Equinox}~\cite{Kidger_arxiv21_Equinox}, and the fPEPS contraction relies on \textsc{quimb}~\cite{Gray_JOSS18_quimb, Gray_PRX24_HyperContraction} and \textsc{symmray}~\cite{Gray_github25_symmray, Gao_PRR25_fTN-contraction}.
In Fig.\,\ref{fig:nn_fpeps}(b), we illustrate the techniques used in our VMC optimization.
All parts of the computational graph have been intensively optimized to achieve high throughput on GPUs, including custom kernels for the HOTRG contraction and for the batched determinants and Pfaffians~\cite{Chen_github26_fermix}.
Several symmetries, including mirror reflections and spin inversion, are imposed by a symmetry projection for better accuracy (SM Sec.\,III~\cite{SM}).
During the VMC optimization, the variational parameters are trained by an adaptive moment stochastic reconfiguration (AdamSR) optimizer, which combines Adam~\cite{Kingma_ICLR15_Adam} with common optimizers for NQS~\cite{Sorella_PRL98_SR, Sorella_PRB01_Lanczos, Chen_NP24_MinSR, Rende_CP24_SRt,Goldshlager_JCP24_SPRING,Gu_NC26_NQS-Hubbard} for better efficiency and stability.
Apart from the common Cholesky solver, we further utilize the iterative least-squares minimal residual (LSMR) linear solver~\cite{Fong_SIAM11_LSMR} in AdamSR to achieve efficient parallelization on up to 1024 GPUs.
The optimizers and solvers are detailed in SM Sec.\,IV~\cite{SM}, and a careful analysis of the time cost is included in SM Sec.\,V~\cite{SM}.
Unless otherwise specified, all simulations are performed with 32-bit floating-point (FP32) precision and TensorFloat-32 (TF32) tensor cores for convolutions and matrix multiplications.

\section{Results}
\label{sec:results}
We now compare the results of the three classes of wavefunctions, namely the BCS backflow state, the Pfaffian backflow state, and the NN-fPEPS, at $1/8$ doping and $U=8$, first at $t' = 0$ and then at $t' = -0.2$.
The variational energies and variances of all our states before and after symmetry projection, together with those of previous studies, are listed in SM Sec.\,VII~\cite{SM}.
The real-space spin, charge, and pair-pair correlation functions of all symmetry-projected states are shown in SM Sec.\,VIII~\cite{SM}.

\begin{figure}[t]
    \centering
    \includegraphics[width=\textwidth]{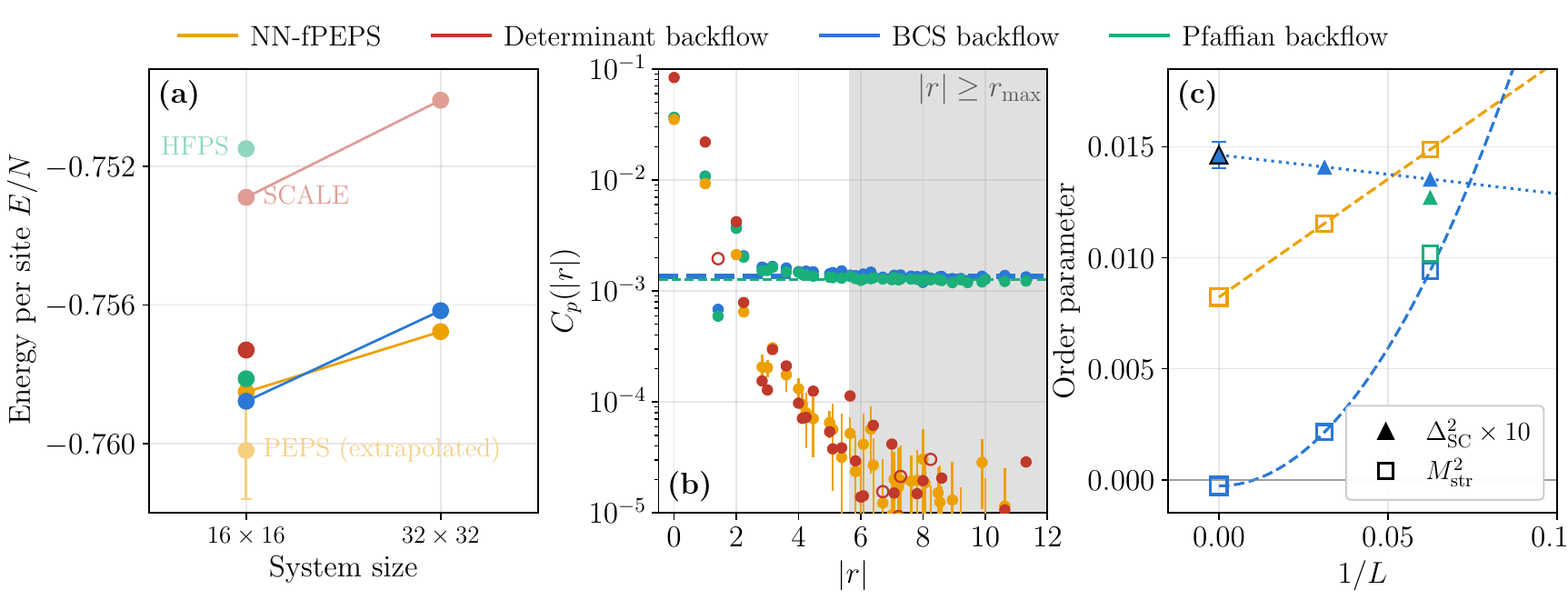}
    \caption{
        Energies and orders of the symmetry-projected states in the pure Hubbard model at $t' = 0$ and $1/8$ doping with PBC.
        (a) Variational energy per site $E/N$ against the system size.
        For reference, we include the variational energies of the determinant backflow ACE (red) and SCALE in Ref.~\cite{Gu_PRX-Intelligence_SCALE}, obtained without symmetry projection, and the hidden fermion Pfaffian state (HFPS) in Ref.~\cite{Roth_arxiv25_HubbardSC}.
        We also show an estimated exact energy obtained by extrapolation of a pure fPEPS calculation with PBC (SM Sec.\,VI~\cite{SM}).
        (b) Pair-pair correlation function $C_p(|r|)$ as defined in Eq.\,\eqref{eq:pairing} on the $16 \times 16$ lattice. For the determinant backflow state, open circles show $|C_p|$ where $C_p < 0$, and the normal-state contribution $\mathcal{N}_{0,r}$ is not subtracted.
        The shaded region is the averaging window $|r| \geq r_{\max} = 4 \sqrt{2}$ of Eq.\,\eqref{eq:delta_SC} and the dashed lines are the resulting $\Delta^2_{\mathrm{SC}}$.
        (c) $d$-wave pairing order parameter $\Delta^2_{\mathrm{SC}}$ as defined in Eq.\,\eqref{eq:delta_SC}, multiplied by $10$, and stripe order parameter $M^2_{\mathrm{str}}$ as defined in Eq.\,\eqref{eq:stripe}, against the inverse linear size $1/L$.
        The dotted and dashed lines are linear fits of $\Delta^2_{\mathrm{SC}} \times 10$ and $M^2_{\mathrm{str}}$ in $1/L$, respectively, and the markers at $1/L = 0$ are the extrapolated values.
        The stripe order of the BCS backflow state is instead fitted in $1/L^2$, as appropriate when there is no long-range order.
        All uncertainties not shown in (c) are below $3 \times 10^{-4}$.
    }
    \label{fig:orders_tp0}
\end{figure}

\subsection{Pure Hubbard model}
\label{sec:pure}
We begin with the pure Hubbard model at $t' = 0$ and $1/8$ doping, where the physics has been more extensively studied than at $t' = -0.2$.
As presented in Fig.\,\ref{fig:orders_tp0}(a), we reach the lowest variational energies reported so far on both the $16 \times 16$ and $32 \times 32$ PBC lattices, below those of the convolutional determinant backflow forms used in ACE and SCALE~\cite{Gu_PRX-Intelligence_SCALE}, of the hidden fermion Pfaffian state (HFPS)~\cite{Roth_arxiv25_HubbardSC}, and of the VAFQMC stripe state~\cite{Sorella_PRB23_VAFQMC, Wu_Science24_Vscore} listed in SM Sec.\,VII~\cite{SM}.
On the $16 \times 16$ lattice, extrapolating a pure fPEPS simple-update calculation with PBC gives an estimated exact energy of $E/N = -0.7602(14)$ (SM Sec.\,VI~\cite{SM}), shown in Fig.\,\ref{fig:orders_tp0}(a).
The energies of all our variational wavefunctions lie within $0.002$~$t$ per site ($0.3\%$) of this estimate, and the lowest one (BCS backflow) lies $0.0014$~$t$ above it. This accuracy is consistent with that of the NN-fPEPS on the small $16 \times 4$ lattice, where an accurate benchmark energy is known (SM Sec.\,VII~\cite{SM}).

In Fig.\,\ref{fig:orders_tp0}(b), the pair-pair correlations of the BCS and Pfaffian backflow states saturate at nearly the same plateau at long distances, $C_p \approx 1.3 \times 10^{-3}$, which signals long-range superconducting correlations.
The NN-fPEPS and the determinant backflow state instead show exponentially decaying correlations, as expected for a pure stripe without pairing, which is also the state found in the large-$D$ fPEPS calculation.
These ansatzes therefore converge to two qualitatively different competing states, a superconductor and a stripe.
As indicated by the energies in Fig.\,\ref{fig:orders_tp0}(a), the NN-fPEPS energy lies below that of the Pfaffian backflow state but slightly above that of the BCS backflow state on the $16 \times 16$ lattice, by $3 \times 10^{-4}$~$t$ per site, while on the $32 \times 32$ lattice its energy is clearly lower than that of the BCS backflow state, by $6 \times 10^{-4}$~$t$ per site.
This trend indicates that the NN-fPEPS captures the striped ground state of large lattices, whereas the superconducting state seen in the BCS and Pfaffian backflow ansatzes is a competing state at higher energy in the TDL.

The order parameters are further analyzed in Fig.\,\ref{fig:orders_tp0}(c) as a function of lattice size.
On the $16 \times 16$ lattice, the BCS backflow state has $\Delta^2_{\mathrm{SC}} = 0.00135(1)$ and $M^2_{\mathrm{str}} = 0.009415(3)$, the weakest stripe order among the three states.
Its pairing strength exceeds even that of the BCS backflow result at $t' = -0.2$ (Sec.\,\ref{sec:ttp}), where superconductivity has been previously reported and argued to be stronger than at $t'=0$.
On the $32 \times 32$ lattice its pairing is essentially unchanged, $\Delta^2_{\mathrm{SC}} = 0.00141(3)$, and extrapolated linearly in $1/L$ it stays finite with $\Delta^2_{\mathrm{SC}} = 0.00146(6)$, i.e., $\Delta_{\mathrm{SC}} = 0.0382(8)$.
Its stripe order, in contrast, drops by a factor of $4.4$ to $M^2_{\mathrm{str}} = 0.002145(4)$, and its extrapolation in $1/L^2$ gives $M^2_{\mathrm{str}} = -0.0003$, consistent with a vanishing stripe order in the TDL.
The BCS backflow wavefunction thus describes a uniform $d$-wave superconductor in the TDL, and its weak stripe signal on finite lattices is a finite-size effect.
Although the Pfaffian backflow state has more variational freedom, it behaves similarly, and its order parameters are comparable to those obtained by the BCS backflow state on the $16 \times 16$ lattice.

The NN-fPEPS, in contrast, gives a filled stripe with $q = \pi/8$, i.e., a spin period of $16$, with the extrapolated stripe order $M^2_{\mathrm{str}} = 0.0082$, while its exponentially decaying pair-pair correlation is consistent with vanishing superconducting order.
The NN-fPEPS and the BCS backflow solutions therefore realize two distinct candidate states in the TDL, a filled stripe without pairing as the ground state and a uniform $d$-wave superconductor as a low-energy competing state.
Our striped ground state agrees with the filled stripes without superconductivity found by a broad range of previous studies at $U = 8$ and $1/8$ doping~\cite{Zheng_Science17_Hubbard, Qin_PRX20_Hubbard, Sorella_PRB23_VAFQMC, Roth_arxiv25_HubbardSC}.
The uniform $d$-wave superconductor expressed by the backflow states has not been commonly seen in recent numerical work, but its low energy matches the energetically competitive uniform $d$-wave state that VAFQMC finds within $0.001$~$t$ per site of the stripe~\cite{Sorella_PRB23_VAFQMC}.
This is a much lower energy scale of competition than seen in early DMET and infinite PEPS studies, which stabilized a uniform $d$-wave state in constrained $2 \times 2$ unit cells at energies of about $0.01$~$t$ per site above the stripe~\cite{Zheng_Science17_Hubbard}.
This indicates the importance of long-range fluctuations in stabilizing the uniform $d$-wave state.
We note that in the closely related $t$--$J$ model, a uniform $d$-wave state has likewise been reported to be among the low-energy states~\cite{Corboz_PRL14_tJ, Yu_arxiv26_tJ}.

\subsection{\texorpdfstring{$t$--$t'$}{t-t'} Hubbard model}
\label{sec:ttp}
\begin{figure}[t]
    \centering
    \includegraphics[width=\textwidth]{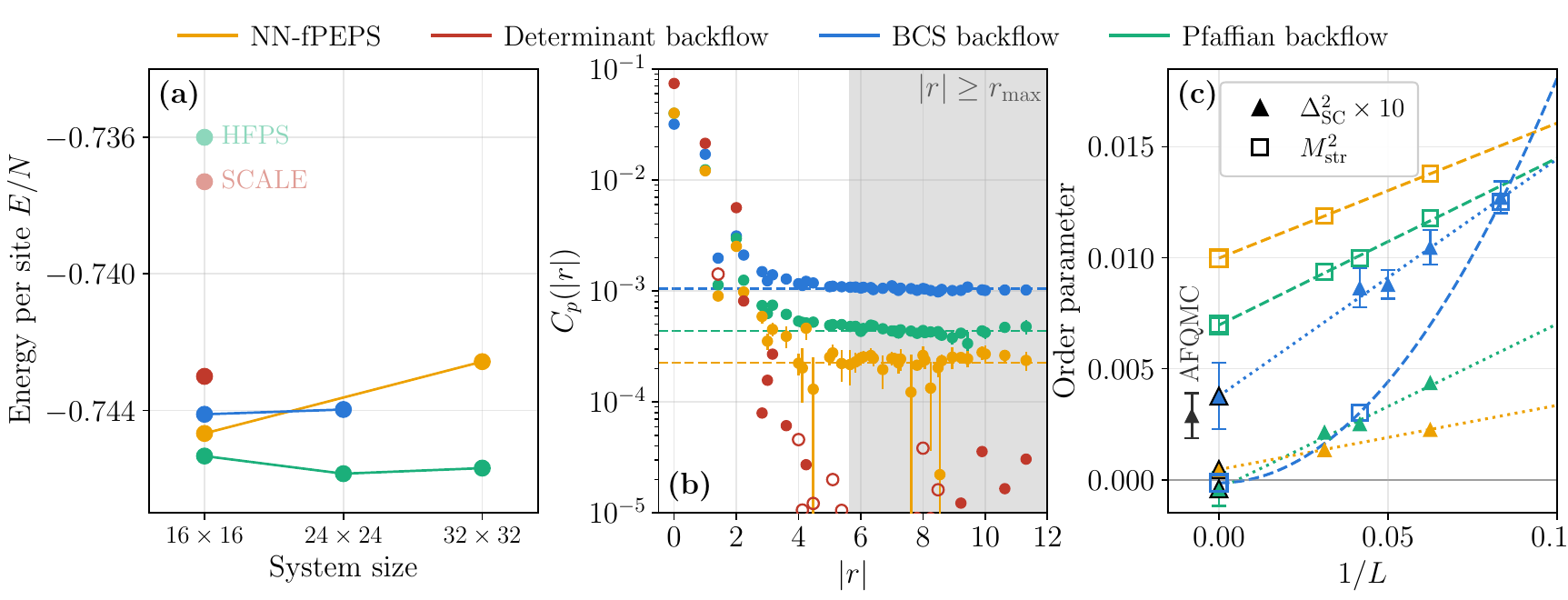}
    \caption{
        Energies and orders of the symmetry-projected states in the $t$--$t'$ Hubbard model, the same as Fig.\,\ref{fig:orders_tp0} but with $t' = -0.2$ ($1/8$ doping).
        The BCS backflow data are extracted from Ref.~\cite{Rende_arxiv26_SBP}.
        In (c), the CP-AFQMC result is from Ref.~\cite{Xu_Science24_Hubbard}, and the stripe order of the BCS backflow state is fitted in $1/L^2$.
    }
    \label{fig:orders_tpm02}
\end{figure}

We now apply a similar analysis to the $t$--$t'$ Hubbard model at $t' = -0.2$ and $1/8$ doping, shown in Fig.\,\ref{fig:orders_tpm02}.
For this model, we take the BCS backflow state energies from the recent study in Ref.~\cite{Rende_arxiv26_SBP} and reanalyze the order parameters of these states.
As presented in Fig.\,\ref{fig:orders_tpm02}(a), our wavefunctions again reach the lowest variational energies reported so far, and the Pfaffian backflow state achieves the lowest energies on all lattice sizes from $16 \times 16$ to $32 \times 32$.
Unlike at $t' = 0$, the pair-pair correlations of the NN-fPEPS, BCS backflow, and Pfaffian backflow states in Fig.\,\ref{fig:orders_tpm02}(b) saturate at a finite plateau, so all three states carry superconducting correlations on finite lattices.
While their short-distance correlations are similar in magnitude, the plateau of the BCS backflow state is $2.4$ times higher than that of the Pfaffian backflow state and $4.6$ times higher than that of the NN-fPEPS.
In contrast, the determinant backflow state has vanishing long-range $d$-wave pairing.
The ansatzes differ even more in the strength of their stripe order, shown in Fig.\,\ref{fig:orders_tpm02}(c).
For the BCS backflow state, the structure factor peaks at $q = \pi/6$ on both the $12 \times 12$ and $24 \times 24$ lattices, where $M^2_{\mathrm{str}} = 0.0125$ and $0.0030$, the latter being already $3.3$ times weaker than the $0.00998(1)$ order of the Pfaffian backflow on the same lattice.
Its peak $S(k_\alpha)$ hardly changes between the two sizes, so the spin correlations are short-ranged, and the extrapolation in $1/L^2$ gives $M^2_{\mathrm{str}} = -0.0001$.
We note that our extrapolation, which starts from the small $12 \times 12$ lattice, is subject to strong finite-size effects and may thus not be entirely reliable.
Nevertheless, since the stripe order is much weaker than that of the other two variational states, we regard the BCS backflow state as evidence of a (nearly) uniform low-energy state in the TDL.
The Pfaffian backflow state and the NN-fPEPS instead form a pronounced stripe order with $q = \pi/4$, i.e., a spin period of $8$, whose $M^2_{\mathrm{str}}$ extrapolates to $0.0100$ for the NN-fPEPS and $0.0070$ for the Pfaffian backflow, indicating a striped state in the TDL.
Similarly, the determinant backflow state also shows a strong stripe order with $q = \pi/4$~\cite{Gu_PRX-Intelligence_SCALE}.
Independent of this consensus between these different ansatzes, the low energy achieved by the Pfaffian backflow wavefunction lends further support to a non-uniform ground state in the TDL.

The extrapolation of the superconducting pairing order is shown in Fig.\,\ref{fig:orders_tpm02}(c).
Extrapolating $\Delta^2_{\mathrm{SC}}$, we find that the BCS backflow state of Ref.~\cite{Rende_arxiv26_SBP} retains a sizable $\Delta_{\mathrm{SC}} = 0.019(4)$ in the TDL, slightly below the value $0.022(1)$ obtained in Ref.~\cite{Rende_arxiv26_SBP} by extrapolating $\Delta_{\mathrm{SC}}$ itself. Since, as discussed above, its stripe order is very weak (and potentially vanishes in a TDL extrapolation), we reinterpret this as evidence of a (nearly) uniform $d$-wave superconducting low-energy state in the TDL, rather than the striped superconducting state originally discussed in Ref.~\cite{Rende_arxiv26_SBP}.
The Pfaffian backflow state and the NN-fPEPS give much weaker pairing in the TDL.
For the NN-fPEPS, the linear fit through the $16 \times 16$ and $32 \times 32$ lattices yields $\Delta_{\mathrm{SC}} = 0.007(2)$.
For the Pfaffian backflow, the result depends on the fit.
The linear $1/L$ fit over $L = 16$, $24$, and $32$ in Fig.\,\ref{fig:orders_tpm02}(c) gives $\Delta_{\mathrm{SC}}$ consistent with zero, with an upper bound of $0.006$ at one standard deviation.
A fit through the $24 \times 24$ and $32 \times 32$ lattices alone gives a small but finite $\Delta_{\mathrm{SC}} = 0.010(3)$.
Either way, the extrapolated pairing gives $\Delta_{\mathrm{SC}} \lesssim 0.010$, i.e., a striped ground state with weak or absent superconductivity.
This stands in contrast to the striped superconducting ground state with a longer wavelength and a larger $\Delta_{\mathrm{SC}} = 0.017(3)$ proposed by CP-AFQMC~\cite{Xu_Science24_Hubbard}.
Taken together, these results thus point to stripes with weaker or absent superconductivity, stripes with robust superconductivity, and a uniform $d$-wave state, all existing in competition near the ground state.

\section{Conclusion and outlook}
\label{sec:conclusion}
We have studied the competition between stripes and superconductivity in the hole-doped Hubbard model with $U=8$ and $1/8$ doping using three classes of variational states: BCS backflow states, Pfaffian backflow states, and NN-fPEPS.
On the lattices considered, from $16 \times 16$ up to $32 \times 32$ sites, the accuracy of our variational energies exceeds that of all prior studies of this model.
In the pure Hubbard model at $t' = 0$, our work lends further support to the long-standing proposal of a non-superconducting striped ground state~\cite{Zheng_Science17_Hubbard,Qin_PRX20_Hubbard}.
At $t'=-0.2$, we find evidence for weak or vanishing superconductivity with strong stripe order in the ground state, which is different from recent studies that saw a robust $d$-wave striped superconducting ground state~\cite{Xu_Science24_Hubbard, Roth_arxiv25_HubbardSC}, although the latter could still represent a different low-lying state~\cite{Zheng_Science17_Hubbard}. Both with and without $t'$, we find at this doping that an ordered stripe suppresses superconductivity, so that the ground state is either not superconducting or only weakly superconducting. However, at both values of $t'$, we also find a strongly superconducting uniform $d$-wave state nearby on a milli-$t$ energy scale, resembling the proposal in Ref.~\cite{Sorella_PRB23_VAFQMC}.

Beyond the physical findings, our work highlights and advances two aspects of current simulation methodology.
First, despite being highly parametrized, modern neural wavefunctions retain biases that are not overcome in training.
They must therefore be cross-validated using different techniques, as done here, in order to identify the qualitative nature of the ground state. Second, the neural tensor networks, improved here from Ref.~\cite{Du_PRB26_NN-fTNS}, now offer state-of-the-art performance in large-scale simulations of a strongly correlated electron system, without using a mean-field reference.
Taken together, these two aspects suggest a new framework where neural wavefunctions with different fermionic references are combined for reliable studies.

We now discuss prospects for future definitive simulations of superconductivity in the Hubbard model ground state. A finite lattice can resolve the pairing only once the gap
$\Delta$ exceeds the mean spacing of the single-particle levels, roughly the bandwidth $8t$ of the square lattice divided by $N$, which requires $N \gtrsim 8t/\Delta$.
For $\Delta_{\mathrm{SC}} \sim 0.02$ comparable to the uniform $d$-wave superconductors in this work and previous estimates at $t' = -0.2$~\cite{Rende_arxiv26_SBP, Xu_Science24_Hubbard}, and taking $\Delta \approx \Delta_{\mathrm{SC}} t$ from BCS theory, eliminating the finite-size effects on the pairing requires $N \gtrsim 400$ sites, a number that is expected to be further reduced in the strong-coupling limit~\cite{Zhang_SST88_RMFT,Paramekanti_PRL01_VMC-Hubbard}.
Running on up to $1024$ GPUs, we simulated $N \approx 1000$ sites and thus have reached a regime where finite-size errors arguably no longer dominate the physics and TDL extrapolation is reliable.
On the other hand, the condensation energy relative to a hypothetical paramagnetic normal state is estimated to be $< 10^{-3}$~$t$~\cite{Gull_PRB12_Energetics}.
If the mechanism driving superconductivity is related to the formation of the striped state, we may expect a competing stripe state to gain some of this energy, and thus to compete with the superconducting state at the milli-$t$ scale or below, which is what we see in our work.
The overall picture is that stripes, striped superconductivity, and uniform superconductivity all exist in extreme competition in the hole-doped Hubbard model, and resolving this competition requires an extreme level of accuracy. Our work presents a concrete step in this direction.

\begin{acknowledgments}
This work was supported by the U.S. Department of Energy, Office of Science, through Award No.~DE-SC0018140.
Most computations were performed using the facilities of the National Energy Research Scientific Computing Center (NERSC), a U.S. Department of Energy Office of Science User Facility located at Lawrence Berkeley National Laboratory, under NERSC Award No.~ERCAP0023924.
This research used resources of NERSC under Contract No. DE-AC02-05CH11231 using NERSC award DDR-ERCAP0038682.
Some computations were also performed using the facilities of the Caltech Resnick High Performance Computing Center.
We acknowledge Yuntian Gu, Christopher Roth, Riccardo Rende, and Antoine Georges for useful discussions about their earlier results.
We thank Yuntian Gu and Yufei Chen for sharing the determinant backflow data of Ref.~\cite{Gu_PRX-Intelligence_SCALE} and Riccardo Rende for sharing the BCS backflow data of Ref.~\cite{Rende_arxiv26_SBP}.
We would like to thank the NVIDIA Corporation for technical support. We are especially grateful to Linsey Rodenbach, Nicola Pancotti, and other members of the NVIDIA Quantum Applied Research team, as well as Katie Klymko (NERSC), for supporting this project.
Claude Code, an AI coding assistant developed by Anthropic, was used to write parts of the simulation code, to submit and monitor part of the computations on the high-performance computing clusters, and to draft parts of this manuscript.
All code, computations, and text were reviewed and verified by the authors, who take full responsibility for the content.
\end{acknowledgments}

\section*{Data and code availability}
The trained wavefunctions, the code to evaluate and measure them, and the data and scripts behind all figures and tables are available at Ref.~\cite{Chen_github26_NeuralHubbard}.

%% file: supplement-setup.tex

\setcounter{figure}{0}
\setcounter{table}{0}
\setcounter{equation}{0}
\setcounter{section}{0}
\renewcommand{\thefigure}{S\arabic{figure}}
\renewcommand{\thetable}{S\arabic{table}}
\renewcommand{\theequation}{S\arabic{equation}}

\providecommand{\theHfigure}{}\renewcommand{\theHfigure}{S\arabic{figure}}
\providecommand{\theHtable}{}\renewcommand{\theHtable}{S\arabic{table}}
\providecommand{\theHequation}{}\renewcommand{\theHequation}{S\arabic{equation}}
\providecommand{\theHsection}{}\renewcommand{\theHsection}{S\arabic{section}}

\renewcommand{\thesection}{\Roman{section}}
\renewcommand{\thesubsection}{\Alph{subsection}}
\setcounter{secnumdepth}{3}

\makeatletter
\def\l@subsubsection#1#2{}
\makeatother

%% file: supplement-body.tex
\begin{center}
\begin{minipage}{0.7\textwidth}
\tableofcontents
\end{minipage}
\end{center}
\vspace{-1pt}

\section{Neural network architecture}
\label{sec:sm_network}

\begin{figure}[tp]
    \centering
    \includegraphics{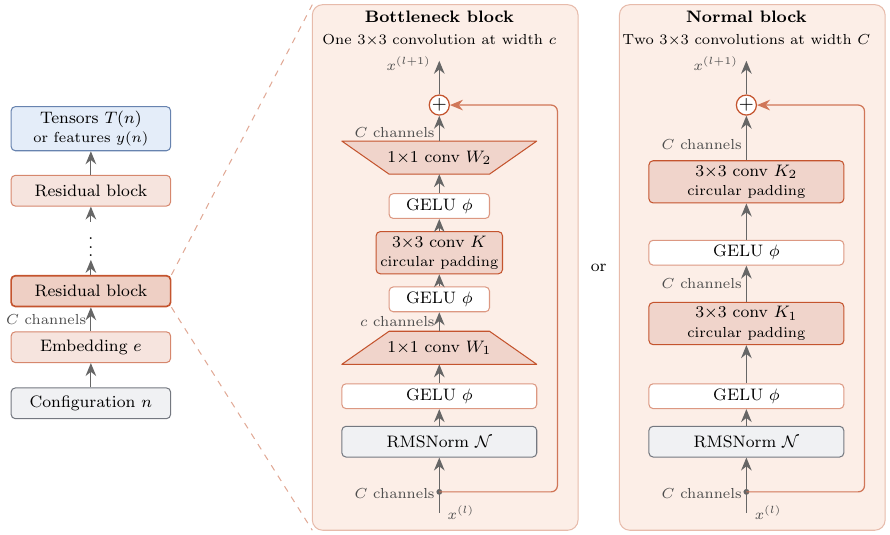}
    \caption{
        Neural network architectures.
        The configuration $n$ is embedded into a $C$-channel feature map and passes through pre-activation residual blocks to yield the site tensors $T(n)$ in the NN-fPEPS or the features $y(n)$ in the backflows.
    }
    \label{fig:sm_network}
\end{figure}

Both the neural-network fermionic projected entangled pair state (NN-fPEPS) and the Bardeen--Cooper--Schrieffer (BCS) and Pfaffian backflow states (here, we use `backflows' for short) generate their configuration-dependent features with a residual convolutional neural network (CNN), sketched in Fig.\,\ref{fig:sm_network} with its settings in Table~\ref{tab:sm_network}.
A learned embedding maps the local state of each site to a $C$-dimensional vector $x_i = e_{n_i}$, with the same table $e$ on every site to preserve translation symmetry, followed by pre-activation residual blocks~\cite{He_CVPR16_ResNet, He_ECCV16_ResNetIdentity}.
The two networks differ in the residual block, which is the bottleneck block in the NN-fPEPS and the normal block in the backflows.
Since the convolutions share their weights across sites, the same network applies to every system size, enabling weight transfer from small to large lattices.

\begin{table}[tp]
    \caption{
        Settings of the two residual CNNs of Fig.\,\ref{fig:sm_network}.
    }
    \label{tab:sm_network}
    \begin{ruledtabular}
    \begin{tabular}{lcc}
        & NN-fPEPS & BCS and Pfaffian backflows \\
        \colrule
        Output per site & $C = 2 D^4$ elements of $T(n)$ & $y^{\uparrow}(n)$ and $y^{\downarrow}(n)$ of dimension $d$ \\
        Residual block & bottleneck & normal \\
        Width $C$ & $512$ & $64$ \\
        Bottleneck width $c$ & $128$ & --- \\
        Weights per block & $2 C c + 9 c^2 \approx 2.8 \times 10^5$ & $2 \times 9 C^2 \approx 7.4 \times 10^4$ \\
        Parameters $N_p$ & $4.5\,\mathrm{M}$ & $1.2\,\mathrm{M}$ \\
        Receptive field & $33 \times 33$ & $65 \times 65$ \\
        \colrule
        Number of blocks & \multicolumn{2}{c}{$16$} \\
        Convolutions & \multicolumn{2}{c}{$3 \times 3$ with circular padding for periodic boundary conditions (PBC), $1 \times 1$ in the bottleneck} \\
        Nonlinearity $\phi$ & \multicolumn{2}{c}{Gaussian error linear unit (GELU)~\cite{Hendrycks_arxiv23_GELU}} \\
        Normalization $\mathcal{N}$ & \multicolumn{2}{c}{RMSNorm~\cite{Zhang_NIPS19_RMSNorm} over all channels and sites, without learnable affine parameters} \\
        Initialization & \multicolumn{2}{c}{LeCun normal~\cite{LeCun_NNTT98_EfficientBackProp}} \\
        Arithmetic & \multicolumn{2}{c}{real-valued, single precision with TensorFloat-32 (TF32) in convolutions and matrix multiplications} \\
    \end{tabular}
    \end{ruledtabular}
\end{table}

The network of the NN-fPEPS maps $n$ to the elements of all local tensors $T^{[i]}(n)$.
A $\mathbb{Z}_2$-graded site tensor has a physical index of dimension $4$, with $\{|0\rangle, |{\uparrow\downarrow}\rangle\}$ even and $\{|{\uparrow}\rangle, |{\downarrow}\rangle\}$ odd, and four virtual indices of dimension $D$, each with $D/2$ even and $D/2$ odd basis states.
Parity conservation removes half of its $4 D^4$ elements, so the network outputs $C = 2 D^4$ real channels per site, $C = 512$ at the bond dimension $D = 4$ used in this work.
Slicing the physical index at $n_i$ gives the tensors contracted by the higher-order tensor renormalization group (HOTRG) of Sec.\,\ref{sec:sm_hotrg}.
This width is fixed by $D$ and far wider than the networks typically used in variational Monte Carlo (VMC), so the NN-fPEPS stacks bottleneck blocks~\cite{He_CVPR16_ResNet}.
A residual block at full width would carry $9 C^2 \approx 2.4 \times 10^6$ weights in its $3 \times 3$ convolution alone, $8.5$ times the parameters and floating-point operations (FLOP) of a bottleneck block.
The width $C = 64$ of the backflows is a free choice rather than being fixed by the bond dimension $D$, so they need no bottleneck.

\clearpage

\section{HOTRG contraction}
\label{sec:sm_hotrg}

\subsection{HOTRG overview}
\label{sec:sm_overview}

For a given configuration $n$, the amplitude $\psi_{\text{NN-fPEPS}}(n)$ is the contraction of a single-layer $L_x \times L_y$ fermionic projected entangled pair state (fPEPS) of four-leg tensors on a torus.
All tensors are $\mathbb{Z}_2$-graded fermionic arrays~\cite{Corboz_PRB10_fPEPS, Kraus_PRA10_fPEPS, Mortier_SP25_fTN}, so every operation acts blockwise on the two fermion-parity sectors of each index and generates the fermionic signs automatically.

\begin{figure}[tp]
    \centering
    \includegraphics{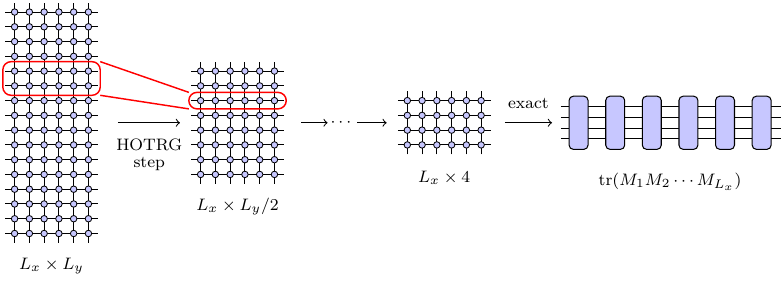}
    \caption{
        The HOTRG contraction scheme of the NN-fPEPS amplitude, illustrated for $L_y = 16$.
        Dangling legs are connected periodically.
        HOTRG steps along the $y$ direction successively halve $L_y$ down to $L_y = 4$, while all bond dimensions are kept no larger than $\chi$.
        The red boxes mark a pair of adjacent rows and the single row of the coarser lattice into which it is merged.
        Each column of the final $L_x \times 4$ network is then contracted into a transfer matrix $M_i$, and the amplitude follows exactly from Eq.\,\eqref{eq:sm_trace}.
    }
    \label{fig:sm_pipeline}
\end{figure}
We contract the network by HOTRG steps~\cite{Xie_PRB12_HOTRG, Iino_PRB19_BoundaryTRG} along $y$ (Fig.\,\ref{fig:sm_pipeline}), a fixed path that suits the regular torus and the batched evaluation of Monte Carlo samples better than the hyperoptimized schemes designed for arbitrary geometries~\cite{Gray_PRX24_HyperContraction}.
The coarse-graining stops at $L_y = 4$, a width that allows for exact contraction at $D = \chi = 4$, and the four tensors of each remaining column are contracted exactly into a transfer matrix $M_i$ of dimension $\chi^4 = 256$.
The exact contraction of these transfer matrices gives
\begin{equation}
    \psi(n) = \mathrm{tr} \left( M_1 M_2 \cdots M_{L_x} \right).
    \label{eq:sm_trace}
\end{equation}
At $D = \chi$ each HOTRG step costs $\mathcal{O}(\chi^7)$ per merged pair and the lattice halves at every step, so the contraction costs $\mathcal{O}(N \chi^7)$ for $N = L_x L_y$ sites.
Because $\chi$ is fixed and no singular-value cutoff is applied, every configuration follows the same sequence of dense linear algebra with static shapes, which is evaluated for a whole Monte Carlo batch in one call on the graphics processing unit (GPU) and keeps the sampled energy strictly variational~\cite{Liu_PRL24_PEPS-VMC}.
The contraction runs in single precision, with TF32 in the final exact contraction, and gradients are obtained by automatic differentiation.

\subsection{HOTRG step}
\label{sec:sm_step}

A step merges every pair of $y$-adjacent tensors $T^{[i,2j]}$ and $T^{[i,2j+1]}$, uniformly across the periodic lattice.
The merged tensor inherits the $x$ legs of both parents, and each doubled $x$ bond is compressed back to dimension $\chi$ by a pair of oblique projectors $P_L P_R \approx \openone$~\cite{Iino_PRB19_BoundaryTRG}.
The projectors on the bond between the coarse sites $(i, j)$ and $(i+1, j)$ are computed from the $2 \times 2$ patch of the two pairs $A_L$ and $A_R$ (dashed boxes), with all bond dimensions set to $\chi$,
\begin{equation}
    \tdg{
        \draw (0,-0.5) -- (0,1.4);
        \draw (1.3,-0.5) -- (1.3,1.4);
        \draw (-0.55,0) -- (1.85,0);
        \draw (-0.55,0.9) -- (1.85,0.9);
        \filldraw[fill=tnsite] (0,0) circle (0.15);
        \filldraw[fill=tnsite] (0,0.9) circle (0.15);
        \filldraw[fill=tnsite] (1.3,0) circle (0.15);
        \filldraw[fill=tnsite] (1.3,0.9) circle (0.15);
        \draw[densely dashed, gray, rounded corners=3pt] (-0.3,-0.32) rectangle (0.3,1.22);
        \draw[densely dashed, gray, rounded corners=3pt] (1.0,-0.32) rectangle (1.6,1.22);
        \node[gray] at (-0.78,0.45) {$A_L$};
        \node[gray] at (2.08,0.45) {$A_R$};
    }
    \;\approx\;
    \tdg{
        \draw (0,-0.55) -- (0,1.45);
        \draw (3.4,-0.55) -- (3.4,1.45);
        \draw (-0.75,0.2) -- (0.85,0.2);
        \draw (-0.75,0.7) -- (0.85,0.7);
        \draw (2.55,0.2) -- (4.15,0.2);
        \draw (2.55,0.7) -- (4.15,0.7);
        \draw (1.45,0.45) -- (1.95,0.45);
        \node[above] at (1.7,0.45) {$\chi$};
        \filldraw[fill=tnproj] (0.85,0.05) -- (0.85,0.85) -- (1.45,0.45) -- cycle;
        \filldraw[fill=tnproj] (2.55,0.05) -- (2.55,0.85) -- (1.95,0.45) -- cycle;
        \node[tblob] at (0,0.45) {$A_L$};
        \node[tblob] at (3.4,0.45) {$A_R$};
        \node at (1.05,-0.25) {$P_L$};
        \node at (2.35,-0.25) {$P_R$};
    }
    \;.
    \label{eq:sm_patch}
\end{equation}
QR and LQ factorizations of $A_L$ and $A_R$, viewed as matrices between their four open legs and the doubled bond, expose the isometries $Q_{L/R}$,
\begin{equation}
    \tdg{
        \draw (-0.7,0.45) -- (0,0.45);
        \draw (-0.7,0.15) -- (0,0.15);
        \draw (-0.7,-0.15) -- (0,-0.15);
        \draw (-0.7,-0.45) -- (0,-0.45);
        \draw (0,0.25) -- (0.7,0.25);
        \draw (0,-0.25) -- (0.7,-0.25);
        \node[tblob] at (0,0) {$A_L$};
    }
    =
    \tdg{
        \draw (-0.45,0.45) -- (0,0.45);
        \draw (-0.45,0.15) -- (0,0.15);
        \draw (-0.45,-0.15) -- (0,-0.15);
        \draw (-0.45,-0.45) -- (0,-0.45);
        \draw (0.55,0.25) -- (1.75,0.25);
        \draw (0.55,-0.25) -- (1.75,-0.25);
        \filldraw[fill=tnfac] (0,-0.55) -- (0,0.55) -- (0.55,0.33) -- (0.55,-0.33) -- cycle;
        \node at (0.23,0) {$Q_L$};
        \node[tbox] at (1.15,0) {$R_L$};
    }
    \,, \qquad
    \tdg{
        \draw (-0.7,0.25) -- (0,0.25);
        \draw (-0.7,-0.25) -- (0,-0.25);
        \draw (0,0.45) -- (0.7,0.45);
        \draw (0,0.15) -- (0.7,0.15);
        \draw (0,-0.15) -- (0.7,-0.15);
        \draw (0,-0.45) -- (0.7,-0.45);
        \node[tblob] at (0,0) {$A_R$};
    }
    =
    \tdg{
        \draw (-0.6,0.25) -- (0.6,0.25);
        \draw (-0.6,-0.25) -- (0.6,-0.25);
        \draw (1.15,0.45) -- (1.6,0.45);
        \draw (1.15,0.15) -- (1.6,0.15);
        \draw (1.15,-0.15) -- (1.6,-0.15);
        \draw (1.15,-0.45) -- (1.6,-0.45);
        \filldraw[fill=tnfac] (0.6,-0.33) -- (0.6,0.33) -- (1.15,0.55) -- (1.15,-0.55) -- cycle;
        \node at (0.92,0) {$Q_R$};
        \node[tbox] at (0,0) {$R_R$};
    }
    \,,
    \label{eq:sm_qr}
\end{equation}
and a truncated singular value decomposition (SVD) of the core $B \equiv R_L R_R$ keeps the largest $\chi/2$ singular values of each parity sector,
\begin{equation}
    \tdg{
        \draw (-0.6,0.25) -- (1.6,0.25);
        \draw (-0.6,-0.25) -- (1.6,-0.25);
        \node[tbox] at (0,0) {$R_L$};
        \node[tbox] at (1.0,0) {$R_R$};
    }
    \;\approx\;
    \tdg{
        \draw (-0.45,0.25) -- (0,0.25);
        \draw (-0.45,-0.25) -- (0,-0.25);
        \draw (0.55,0) -- (1.85,0);
        \draw (2.4,0.25) -- (2.85,0.25);
        \draw (2.4,-0.25) -- (2.85,-0.25);
        \filldraw[fill=tnfac] (0,-0.4) -- (0,0.4) -- (0.55,0) -- cycle;
        \filldraw[fill=tnfac] (2.4,-0.4) -- (2.4,0.4) -- (1.85,0) -- cycle;
        \node[tdia] at (1.2,0) {};
        \node at (0.14,-0.6) {$U_\chi$};
        \node at (1.2,-0.6) {$S_\chi$};
        \node at (2.26,-0.6) {$V_\chi^\dagger$};
        \path (0,-0.78) (0,0.78);
    }
    \,,
    \label{eq:sm_svd}
\end{equation}
so that $Q_L U_\chi S_\chi V_\chi^\dagger Q_R$ is the optimal rank-$\chi$ approximation of the patch.
Requiring $A_L P_L = Q_L U_\chi S_\chi^{1/2}$ and $P_R A_R = S_\chi^{1/2} V_\chi^\dagger Q_R$ gives the projectors
\begin{equation}
    \tdg{
        \draw (-0.45,0.25) -- (0,0.25);
        \draw (-0.45,-0.25) -- (0,-0.25);
        \draw (0.55,0) -- (1.0,0);
        \filldraw[fill=tnproj] (0,-0.4) -- (0,0.4) -- (0.55,0) -- cycle;
        \node at (0.16,-0.6) {$P_L$};
        \path (0,-0.78) (0,0.78);
    }
    =
    \tdg{
        \draw (-0.6,0.25) -- (1.0,0.25);
        \draw (-0.6,-0.25) -- (1.0,-0.25);
        \draw (1.55,0) -- (2.7,0);
        \node[tbox] at (0,0) {$R_L^{-1}$};
        \filldraw[fill=tnfac] (1.0,-0.4) -- (1.0,0.4) -- (1.55,0) -- cycle;
        \node[tdia] at (2.15,0) {};
        \node at (1.16,-0.6) {$U_\chi$};
        \node at (2.15,-0.6) {$S_\chi^{1/2}$};
        \path (0,-0.78) (0,0.78);
    }
    \,, \qquad
    \tdg{
        \draw (-0.45,0) -- (0,0);
        \draw (0.55,0.25) -- (1.0,0.25);
        \draw (0.55,-0.25) -- (1.0,-0.25);
        \filldraw[fill=tnproj] (0.55,-0.4) -- (0.55,0.4) -- (0,0) -- cycle;
        \node at (0.4,-0.6) {$P_R$};
        \path (0,-0.78) (0,0.78);
    }
    =
    \tdg{
        \draw (-0.45,0) -- (1.15,0);
        \draw (1.7,0.25) -- (3.05,0.25);
        \draw (1.7,-0.25) -- (3.05,-0.25);
        \node[tdia] at (0,0) {};
        \filldraw[fill=tnfac] (1.7,-0.4) -- (1.7,0.4) -- (1.15,0) -- cycle;
        \node[tbox] at (2.45,0) {$R_R^{-1}$};
        \node at (0,-0.6) {$S_\chi^{1/2}$};
        \node at (1.55,-0.6) {$V_\chi^\dagger$};
        \path (0,-0.78) (0,0.78);
    }
    \,,
    \label{eq:sm_proj}
\end{equation}
which satisfy $P_L P_R = \openone$ without truncation.
Each pair $A^{[i,j]}$ is then contracted with its two neighboring projectors into the coarse tensor $T'^{[i,j]}$,
\begin{equation}
    \tdg{
        \draw (0,-0.5) -- (0,1.4);
        \draw (-0.75,0) -- (0.75,0);
        \draw (-0.75,0.9) -- (0.75,0.9);
        \filldraw[fill=tnsite] (0,0) circle (0.17);
        \filldraw[fill=tnsite] (0,0.9) circle (0.17);
        \draw[densely dashed, gray, rounded corners=3pt] (-0.3,-0.32) rectangle (0.3,1.22);
        \node[gray] at (0,1.62) {$A^{[i,j]}$};
        \node at (-1.05,0.45) {$\chi^2$};
    }
    \;\approx\;
    \tdg{
        \draw (0,-0.5) -- (0,1.4);
        \draw (-0.7,0) -- (0.7,0);
        \draw (-0.7,0.9) -- (0.7,0.9);
        \draw (-1.85,0.45) -- (-1.32,0.45);
        \draw (1.32,0.45) -- (1.85,0.45);
        \filldraw[fill=tnproj] (-0.7,-0.12) -- (-0.7,1.02) -- (-1.32,0.45) -- cycle;
        \filldraw[fill=tnproj] (0.7,-0.12) -- (0.7,1.02) -- (1.32,0.45) -- cycle;
        \filldraw[fill=tnsite] (0,0) circle (0.17);
        \filldraw[fill=tnsite] (0,0.9) circle (0.17);
        \node at (-1.1,-0.4) {$P_R^{[i-1,j]}$};
        \node at (1.05,-0.4) {$P_L^{[i,j]}$};
    }
    \;=\;
    \tdg{
        \draw (0,-0.5) -- (0,1.4);
        \draw (-0.75,0.45) -- (0.75,0.45);
        \filldraw[fill=tnsite] (0,0.45) circle (0.27);
        \node at (0,0.45) {$T'$};
        \node at (0.95,0.45) {$\chi$};
    }
    \,,
    \label{eq:sm_merge}
\end{equation}
where $P_{L/R}^{[i,j]}$ are given by Eqs.\,\eqref{eq:sm_patch}--\eqref{eq:sm_proj} with $A_L = A^{[i,j]}$ and $A_R = A^{[i+1,j]}$.

The complexity of each step is summarized in Table~\ref{tab:sm_hotrg}.

\subsection{Implementation}

A direct evaluation of Eqs.\,\eqref{eq:sm_qr}--\eqref{eq:sm_proj} is not optimal on GPUs.
The triangular inverses in Eq.\,\eqref{eq:sm_proj} are ill-conditioned, and dense QR and SVD routines parallelize poorly and complicate automatic differentiation.
Following \textsc{quimb}~\cite{Gray_JOSS18_quimb}, we therefore replace every stage by matrix products, Cholesky factorizations, and Hermitian eigendecompositions, as summarized in Table~\ref{tab:sm_hotrg} and described below in the order of the algorithm.
All matrices are block diagonal in fermion parity, so each factorization acts on two blocks of half the size, batched together with the doubled bonds and the Monte Carlo samples in a single call.
At $D = \chi = 4$ these blocks are only $8 \times 8$, a size at which generic fermionic tensor libraries~\cite{Gray_github25_symmray, Gao_PRR25_fTN-contraction} are bound by memory latency rather than arithmetic.
The performance-critical stages therefore run as custom fused GPU kernels, each passing over the tensor data once instead of streaming it through a long chain of small library calls.
These kernels were prepared with the assistance of Claude Code, as acknowledged in the main text.

The first stage computes the triangular factors $R_{L/R}$ of Eq.\,\eqref{eq:sm_qr}.
The projectors need only these factors and never the isometries $Q_{L/R}$, so we use the faster Cholesky QR algorithm instead of a full QR factorization.
The factors follow from the Gram matrices $M_L = A_L^\dagger A_L = R_L^\dagger R_L$ and $M_R = A_R A_R^\dagger = R_R R_R^\dagger$, which are contracted directly as tensor networks of each pair with its own conjugate, so that $A_{L/R}$ is never formed explicitly.
Forming and factorizing the Gram matrices costs $\mathcal{O}(\chi^6)$, compared with $\mathcal{O}(\chi^8)$ for a dense QR factorization of $A_{L/R}$.
The Cholesky factorization requires a positive-definite input, whereas $M_{L/R}$ is only positive semidefinite and has the squared condition number of $A_{L/R}$.
We therefore regularize it by the relative diagonal shift
\begin{equation}
    M \to M + \epsilon \, \mathrm{tr}(M) \, \openone,
    \label{eq:sm_shift}
\end{equation}
with $\epsilon = 10^{-4}$.

The second stage computes the truncated SVD of the core $B = R_L R_R$ in Eq.\,\eqref{eq:sm_svd}.
As the SVD is slow on GPUs, we use the Hermitian eigendecomposition of the Gram matrix of $B$,
\begin{equation}
    B^\dagger B = V \Lambda V^\dagger, \qquad
    S_\chi = \Lambda_\chi^{1/2}, \qquad
    U_\chi = B V_\chi S_\chi^{-1},
    \label{eq:sm_eig}
\end{equation}
where $\Lambda_\chi$ and $V_\chi$ contain the $\chi/2$ largest eigenpairs of each parity sector.
The Gram matrix $B^\dagger B$ is regularized by Eq.\,\eqref{eq:sm_shift} with $\epsilon = 10^{-6}$, which keeps the retained singular values away from zero, so that the inverse powers of $S_\chi$ here and in Eq.\,\eqref{eq:sm_proj_fast} stay well conditioned.

The third stage assembles the projectors.
Inserting $R_L^{-1} = R_R B^{-1}$, $R_R^{-1} = B^{-1} R_L$, and $B^{-1} = V S^{-1} U^\dagger$ into Eq.\,\eqref{eq:sm_proj} gives the robust projectors
\begin{equation}
    P_L = R_R V_\chi S_\chi^{-1/2}, \qquad
    P_R = S_\chi^{-1/2} U_\chi^\dagger R_L.
    \label{eq:sm_proj_fast}
\end{equation}
They are mathematically equivalent to Eq.\,\eqref{eq:sm_proj}, but the triangular inverses are replaced by matrix products.
The only remaining inverse, $S_\chi^{-1/2}$, involves only the retained singular values, which the shift above keeps bounded away from zero.

The fourth stage merges each pair $A^{[i,j]}$ with its two neighboring projectors as in Eq.\,\eqref{eq:sm_merge} with a dominant cost of $\mathcal{O}(\chi^7)$, and it runs as a fused kernel.

After the last HOTRG step, the final contraction evaluates Eq.\,\eqref{eq:sm_trace}.
A fused kernel builds the transfer matrices $M_i$ directly in their parity-blocked layout.
Finally, the VMC optimization differentiates $\psi(n)$ through the entire contraction.
The reverse-mode gradient of the eigendecomposition in Eq.\,\eqref{eq:sm_eig} contains denominators $1/(\lambda_k - \lambda_l)$ that diverge for nearly degenerate eigenvalues.
We replace them by the broadened form $(\lambda_k - \lambda_l) / [(\lambda_k - \lambda_l)^2 + \varepsilon^2]$ with a small $\varepsilon$ proportional to the spectral radius~\cite{Liao_PRX19_DiffTN}, which keeps the gradients stable throughout the optimization.

\begin{table}[tp]
    \caption{
        Implementation of the HOTRG contraction.
        Each row replaces the direct evaluation of a stage by GPU-friendly operations, with the costs at $D = \chi$.
        The fused kernels evaluate the whole contraction $3$ to $4$ times faster than a direct implementation.
    }
    \label{tab:sm_hotrg}
    \begin{ruledtabular}
    \begin{tabular}{llll}
        Stage & Direct evaluation & Implementation & Cost \\
        \colrule
        Factors $R_{L/R}$ & QR and LQ of $A_{L/R}$, $\mathcal{O}(\chi^8)$ & Cholesky of $M_{L/R}$, $\epsilon = 10^{-4}$ & $\mathcal{O}(\chi^6)$ \\
        Truncated SVD & SVD of $B$, Eq.\,\eqref{eq:sm_svd} & eigendecomposition, Eq.\,\eqref{eq:sm_eig}, $\epsilon = 10^{-6}$ & $\mathcal{O}(\chi^6)$ \\
        Projectors & $R_{L/R}^{-1}$, Eq.\,\eqref{eq:sm_proj} & matrix products, Eq.\,\eqref{eq:sm_proj_fast} & $\mathcal{O}(\chi^5)$ \\
        Merge & Eq.\,\eqref{eq:sm_merge} & fused kernel & $\mathcal{O}(\chi^7)$ \\
        Final contraction & Eq.\,\eqref{eq:sm_trace} & fused kernel & --- \\
        Gradient of Eq.\,\eqref{eq:sm_eig} & $1/(\lambda_k - \lambda_l)$ & $(\lambda_k - \lambda_l) / [(\lambda_k - \lambda_l)^2 + \varepsilon^2]$~\cite{Liao_PRX19_DiffTN} & --- \\
    \end{tabular}
    \end{ruledtabular}
\end{table}

\section{Symmetry}
\label{sec:sm_symm}

\subsection{Symmetry in fermion systems}
\label{sec:sm_fermion_symm}

Projecting a variational state onto an irreducible representation of the symmetry group $G$ of the Hamiltonian typically lowers its variational energy~\cite{Tahara_JPSJ08_mVMC, Nomura_JPCM2021_RBMsymm}, but for fermions the permutation of modes generates a sign that has to be tracked explicitly~\cite{Misawa_CPC19_mVMC, Chen_PNAS26_HFPS}.
A symmetry $g \in G$ permutes the single-particle modes $x$, comprising site and spin, as $\hat{g} \, c^\dagger_x \, \hat{g}^{-1} = c^\dagger_{g x}$ with $\hat{g} |0\rangle = |0\rangle$.
On the Fock state $|n\rangle = c^\dagger_{x_1} \cdots c^\dagger_{x_{N_e}} |0\rangle$ with $x_1 < \cdots < x_{N_e}$ in a fixed reference order, it acts as
\begin{equation}
    \hat{g} |n\rangle = c^\dagger_{g x_1} \cdots c^\dagger_{g x_{N_e}} |0\rangle = \Pi(n, g) \, |g n\rangle \,,
    \label{eq:sm_symm_fock}
\end{equation}
where $\Pi(n, g) = \pm 1$ is the signature of the permutation that sorts $g x_1, \ldots, g x_{N_e}$ back into the reference order.
On a spinless four-site ring, for instance, $\hat{\mathcal{T}}_1 c^\dagger_2 c^\dagger_4 |0\rangle = c^\dagger_3 c^\dagger_1 |0\rangle = - |1, 0, 1, 0\rangle$, where $\Pi(n, g) = -1$.
The projection $|\Psi^{\text{symm}}\rangle = |G|^{-1} \sum_{g} \omega^*_g \, \hat{g} |\Psi\rangle$ onto the representation with characters $\omega_g$ then has the amplitudes
\begin{equation}
    \psi^{\text{symm}}(n) = \frac{1}{|G|} \sum_{g \in G} \Pi(n, g) \, \omega_g \, \psi(g n) \,.
    \label{eq:sm_symm_amp}
\end{equation}
We project onto the trivial representation, $\omega_g = 1$, of the group generated by the $N$ translations $\mathcal{T}_t$ and
\begin{equation}
    P = \{1, M_x\} \times \{1, M_y\} \times \{1, Z\} \,,
    \label{eq:sm_symm_pointgroup}
\end{equation}
where $M_x: (x, y) \mapsto (-x, y)$ and $M_y: (x, y) \mapsto (x, -y)$ are the mirrors about the lattice axes and $Z$ exchanges the two spin species.
Each element of $P$ costs one more network evaluation and contraction, a size-independent factor of $8$.

\subsection{Translation symmetry in NN-fPEPS}
\label{sec:sm_hotrg_trans}

\begin{figure}[tp]
    \centering
    \includegraphics{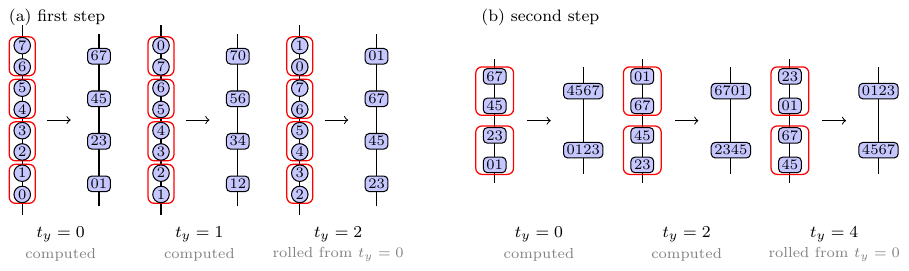}
    \caption{
        Sharing of HOTRG steps among translations, illustrated on one column of $L_y = 8$ sites.
        Circles are the site tensors of the translated configuration $\mathcal{T}_t n$, labeled by their original site indices, red boxes mark the fixed pairs merged by the step, and each coarse tensor is labeled by the original sites it contains.
        (a) In the first step, $t_y = 2$ merges the same pairs of tensors as $t_y = 0$, so its output chain is a cyclic roll of the $t_y = 0$ output and requires no new computation, whereas $t_y = 1$ merges new pairs.
        (b) In the second step, which acts on the chains produced by the first step, $t_y = 4$ is a roll of $t_y = 0$, whereas $t_y = 2$ must be computed.
        In general, two translations whose $t_y$ differ by a multiple of $2^k$ enter the $k$-th step as rolls of each other, so this step has to be computed only for $2^k$ representative translations, one for each value of $t_y \bmod 2^k$, and all others follow by rolling.
    }
    \label{fig:sm_share}
\end{figure}

The circular convolutions make the network translation covariant,
\begin{equation}
    T^{[r]}(\mathcal{T}_t n) = T^{[r + t]}(n),
    \label{eq:sm_covariance}
\end{equation}
so an exact contraction would give translation-symmetric amplitudes.
HOTRG merges the fixed pairs $(2j, 2j+1)$, however, so configurations translated along $y$ are truncated differently, and $\psi$ of Eq.\,\eqref{eq:sm_trace} keeps only the translation symmetry along $x$.
We restore it by the zero-momentum projection
\begin{equation}
    \psi^{\text{trans}}(n) = \frac{1}{N} \sum_{t} \Pi(n, \mathcal{T}_t) \, \psi(\mathcal{T}_t n),
    \label{eq:sm_transavg}
\end{equation}
which naively costs $N$ contractions, $\mathcal{O}(N^2 \chi^7)$ per configuration.
Instead, the network runs once, since by Eq.\,\eqref{eq:sm_covariance} all translated tensor grids are rolls of the same grid, and the HOTRG steps are shared among the translations, as presented in Fig.\,\ref{fig:sm_share}.
The $k$-th step is evaluated for $2^k$ representatives on a lattice of $N / 2^k$ tensors, so that
\begin{equation}
    \sum_{k = 1}^{\log_2 (L_y/4)}
    \underbrace{2^k}_{\text{representatives}} \times
    \underbrace{\mathcal{O}\!\left( N \chi^7 / 2^k \right)}_{\text{cost of one step}}
    = \mathcal{O}\!\left( N \chi^7 \log N \right),
    \label{eq:sm_symm_cost}
\end{equation}
and the final exact contraction, performed once per representative, does not change this scaling.

The group $P$ in Eq.\,\eqref{eq:sm_symm_pointgroup} is then applied as an outer loop,
\begin{equation}
    \psi^{\text{symm}}(n) = \frac{1}{|P|} \sum_{p \in P} \Pi(n, p) \, \psi^{\text{trans}}(p \, n) \,,
    \label{eq:sm_pointavg}
\end{equation}
which realizes Eq.\,\eqref{eq:sm_symm_amp} for the full group.
All results are obtained with $\psi^{\text{symm}}(n)$ unless otherwise specified.

\section{Optimizer}
\label{sec:sm_optimizer}

\subsection{AdamSR}
\label{sec:sm_adamsr}

\begin{algorithm}[t]
    \caption{AdamSR}
    \label{alg:sm_adamsr}
    \begin{algorithmic}[1]
        \Require momentum factors $\mu = 0.95$ and $\beta = 0.995$, norm-constrained step size $\gamma_t$ of Eq.\,\eqref{eq:sm_clip} or Eq.\,\eqref{eq:sm_constraint}
        \State $m_0 \gets 0$, \quad $v_0 \gets 0$
        \For{$t = 1, 2, \dots$}
            \State sample configurations $n_s$ from $|\psi(n)|^2$ and construct $\bar{O}$ and $\bar{\epsilon}$ in Eq.\,\eqref{eq:sm_sreq}
            \State $g_t \gets \bar{O}^+ \bar{\epsilon}$ \label{algline:sm_g} \Comment{plain MinSR solution}
            \State $m_t \gets \mu\, m_{t-1} + (1 - \mu)\, g_t$, \quad $\hat{m}_t \gets m_t / \left(1 - \mu^t\right)$ \label{algline:sm_m} \Comment{Adam first moment}
            \State $v_t \gets \beta\, v_{t-1} + (1 - \beta)\, |g_t|^2$, \quad $\hat{v}_t \gets v_t / \left(1 - \beta^t\right)$ \label{algline:sm_v} \Comment{Adam second moment}
            \State $V_t \gets \mathrm{diag}\big(\hat{v}_t\big)^{1/4} + 10^{-8} \, \openone$ \label{algline:sm_V} \Comment{diagonal preconditioner}
            \State $\delta\theta_t \gets \hat{m}_t + V_t^{-1} \big( \bar{O} V_t^{-1} \big)^+ \big( \bar{\epsilon} - \bar{O} \hat{m}_t \big)$ \label{algline:sm_step} \Comment{momentum $+$ preconditioned correction}
            \State $\theta_t \gets \theta_{t-1} - \gamma_t \big( \lVert \delta\theta_t \rVert \big) \, \delta\theta_t$ \label{algline:sm_update} \Comment{norm-constrained update}
        \EndFor
    \end{algorithmic}
\end{algorithm}

\begin{figure}[b]
    \includegraphics[width=0.6\textwidth]{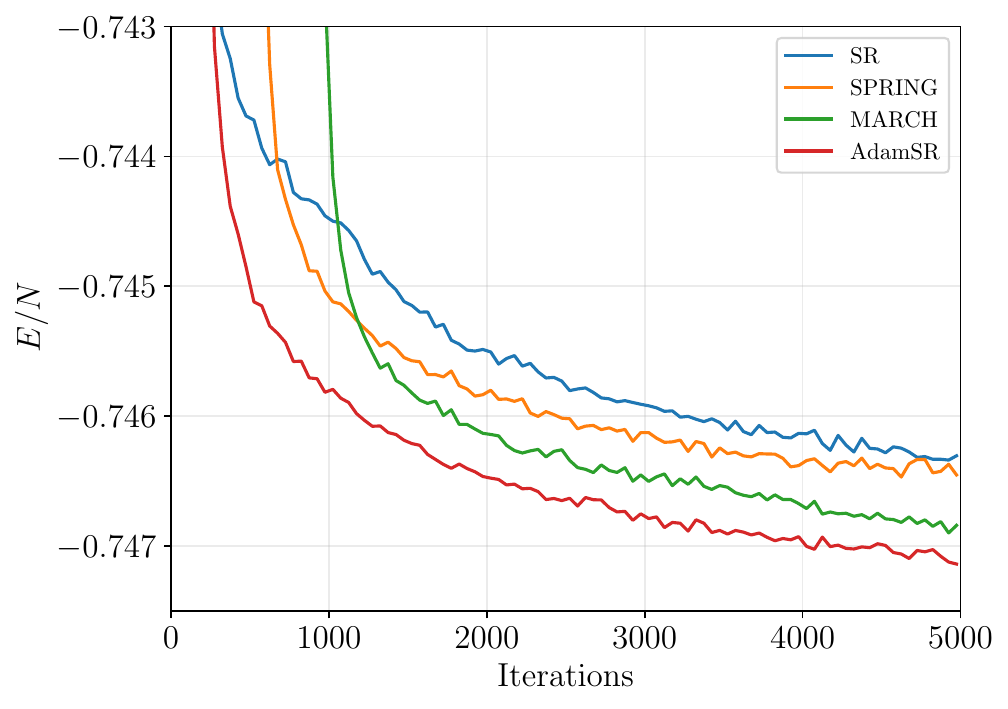}
    \caption{
        Benchmark of optimizers of the stochastic reconfiguration (SR) family on the $8 \times 8$ Hubbard model at $U = 8$, $t' = -0.2$, and $1/8$ hole doping, using the BCS backflow.
        Each curve is the best run over the scanned $C_0$ and Monte Carlo sweep lengths, $C_0 = 1$ for SR and AdamSR and $C_0 = 0.5$ for SPRING and MARCH, averaged over bins of 50 iterations.
    }
    \label{fig:sm_optbench}
\end{figure}

The variational parameters $\theta$ are optimized by stochastic reconfiguration (SR)~\cite{Sorella_PRL98_SR, Sorella_PRB01_Lanczos}, which realizes imaginary-time evolution projected onto the variational manifold.
At every iteration, $N_s$ configurations $n_s$ are sampled from $|\psi(n)|^2$ by Markov-chain Monte Carlo (MCMC), and the update $\dot\theta$ of the $N_p$ parameters solves
\begin{equation}
    \bar{O} \, \dot\theta = -\bar{\epsilon} \,,
    \qquad
    \bar{O}_{sk} = \frac{O_k(n_s) - \left< O_k \right>}{\sqrt{N_s}} \,,
    \qquad
    \bar{\epsilon}_s = \frac{E_{\mathrm{loc}}(n_s) - \left< E_{\mathrm{loc}} \right>}{\sqrt{N_s}} \,,
    \label{eq:sm_sreq}
\end{equation}
with $O_k(n) = \partial \ln \psi(n) / \partial \theta_k$, the local energy $E_{\mathrm{loc}}(n) = \sum_{n'} \left< n|H|n' \right> \psi(n') / \psi(n)$, and $\left< \, \cdot \, \right>$ the sample average.

Since $N_s \ll N_p$, Eq.\,\eqref{eq:sm_sreq} is underdetermined and has infinitely many solutions.
This freedom lets the solver carry the first and second moments of the update, which accelerate and stabilize first-order optimizers such as Adam~\cite{Kingma_ICLR15_Adam}, without giving up the SR equation.
In previous works, SPRING~\cite{Goldshlager_JCP24_SPRING} incorporates the first moment, and MARCH~\cite{Gu_NC26_NQS-Hubbard} further includes the second moment.
In this work, we introduce the adaptive moment stochastic reconfiguration (AdamSR) as presented in Algorithm~\ref{alg:sm_adamsr}, which keeps both ingredients but estimates the moments more robustly.
The price is a second pseudoinverse per iteration, which roughly doubles the solver cost of plain SR.

The norm-constrained step size $\gamma_t$, also following MARCH, preserves the update direction and either clips the norm of $\delta\theta_t$ at $C_0$,
\begin{equation}
    \gamma_t \big( \lVert \delta\theta_t \rVert \big) = \gamma \, \min \left( 1, \frac{C_0}{\lVert \delta\theta_t \rVert} \right) \,,
    \label{eq:sm_clip}
\end{equation}
or regulates it to a scheduled target $c_t$,
\begin{equation}
    \gamma_t \big( \lVert \delta\theta_t \rVert \big) = \min \left( 1, \frac{c_t}{\lVert \delta\theta_t \rVert} \right) \,, \qquad c_t = \frac{C_0}{1 + \max(t - t_0, 0) / T} \,,
    \label{eq:sm_constraint}
\end{equation}
which the raw step norm typically exceeds, so that the realized step norm is $c_t$ itself.
We use Eq.\,\eqref{eq:sm_clip} for the NN-fPEPS and Eq.\,\eqref{eq:sm_constraint} for the backflows, and observe no significant difference between the two.
In the benchmark of Fig.\,\ref{fig:sm_optbench} performed on the BCS backflow, AdamSR tolerates a larger $C_0$ than SPRING and MARCH and converges fastest.
Its advantage is even clearer for the NN-fPEPS, which SR, SPRING, and MARCH fail to train to good convergence in our numerical experiments.

\subsection{Solving linear equations}
\label{sec:sm_solver}

Lines~\ref{algline:sm_g} and \ref{algline:sm_step} of Algorithm~\ref{alg:sm_adamsr} solve
\begin{equation}
    A x = b \,, \qquad x = A^+ b \,,
    \label{eq:sm_axb}
\end{equation}
where $A$ has the $N_s \times N_p$ shape of $\bar{O}$, is distributed across $N_d$ devices along the sample axis, and is given in single precision together with $b$.
Since $N_s \ll N_p$, Eq.\,\eqref{eq:sm_axb} is consistent and massively underdetermined.
We compare three solvers.
The Cholesky solver evaluates the minimum-step SR (MinSR) solution~\cite{Chen_NP24_MinSR,Rende_CP24_SRt}, which inverts the kernel in sample space by a Cholesky factorization,
\begin{equation}
    x = A^\dagger \left( A A^\dagger + \lambda \openone \right)^{-1} b \,.
    \label{eq:sm_minsr}
\end{equation}
The conjugate gradient (CG) method~\cite{Shewchuk_94_cg, Neuscamman_PRB12_SR, Vicentini_SciPost22_netket3} solves the normal equation
\begin{equation}
    A^\dagger A \, x = A^\dagger b
    \label{eq:sm_plainsr}
\end{equation}
through the products $A^\dagger (A v)$.
The least-squares minimal residual (LSMR) method~\cite{Fong_SIAM11_LSMR}, as implemented in \textsc{lineax}~\cite{Rader_arxiv23_Lineax}, is a Krylov iteration on Eq.\,\eqref{eq:sm_axb} itself through the Golub--Kahan bidiagonalization of $A$ (which reduces $A$ to bidiagonal form, i.e., nonzero only on the diagonal and first subdiagonal).
Neither Krylov solver carries a diagonal shift.
Starting from $x = 0$, both approach the minimum-norm solution within a Krylov space that grows by one direction per iteration, so the stopping criterion alone sets the accuracy and acts as the regularizer~\cite{Hansen_SIAM98_IllPosed}.

The Cholesky solver relies on a matrix factorization, whereas CG and LSMR are iterative, and the two classes differ in three respects.
\begin{itemize}
    \item \emph{Precision.} The Cholesky solver forms $A A^\dagger$, whose condition number is the square of that of $A$, and therefore needs double precision (FP64). Both Krylov solvers run in single precision (FP32), where the error floor is set by accumulating the matrix--vector products over $N_p$ terms, $\epsilon_{32} \sqrt{N_p} \sim 10^{-4}$ for the FP32 machine epsilon $\epsilon_{32} \approx 1.2 \times 10^{-7}$, rather than by the squared condition number. This also suits hardware with reduced or no double-precision throughput, such as the NVIDIA B300 GPU and Google TPUs.
    \item \emph{Parallelism.} A Krylov iteration accesses $A$ only through $A v$ and $A^\dagger u$, so it distributes over many nodes with a fixed memory footprint per device. The factorization needs the whole $N_s \times N_s$ kernel, which is either replicated on every device or factorized across devices at a substantial communication cost, as in \textsc{JAXMg}~\cite{Tutt_github26_JAXMg}.
    \item \emph{Accuracy.} The cost of a factorization does not depend on the shift, so the Cholesky solver reaches an arbitrarily small residual simply by lowering $\lambda$. A Krylov solver can only approach its target by adding iterations.
\end{itemize}

\begin{figure}[tp]
    \includegraphics[width=\textwidth]{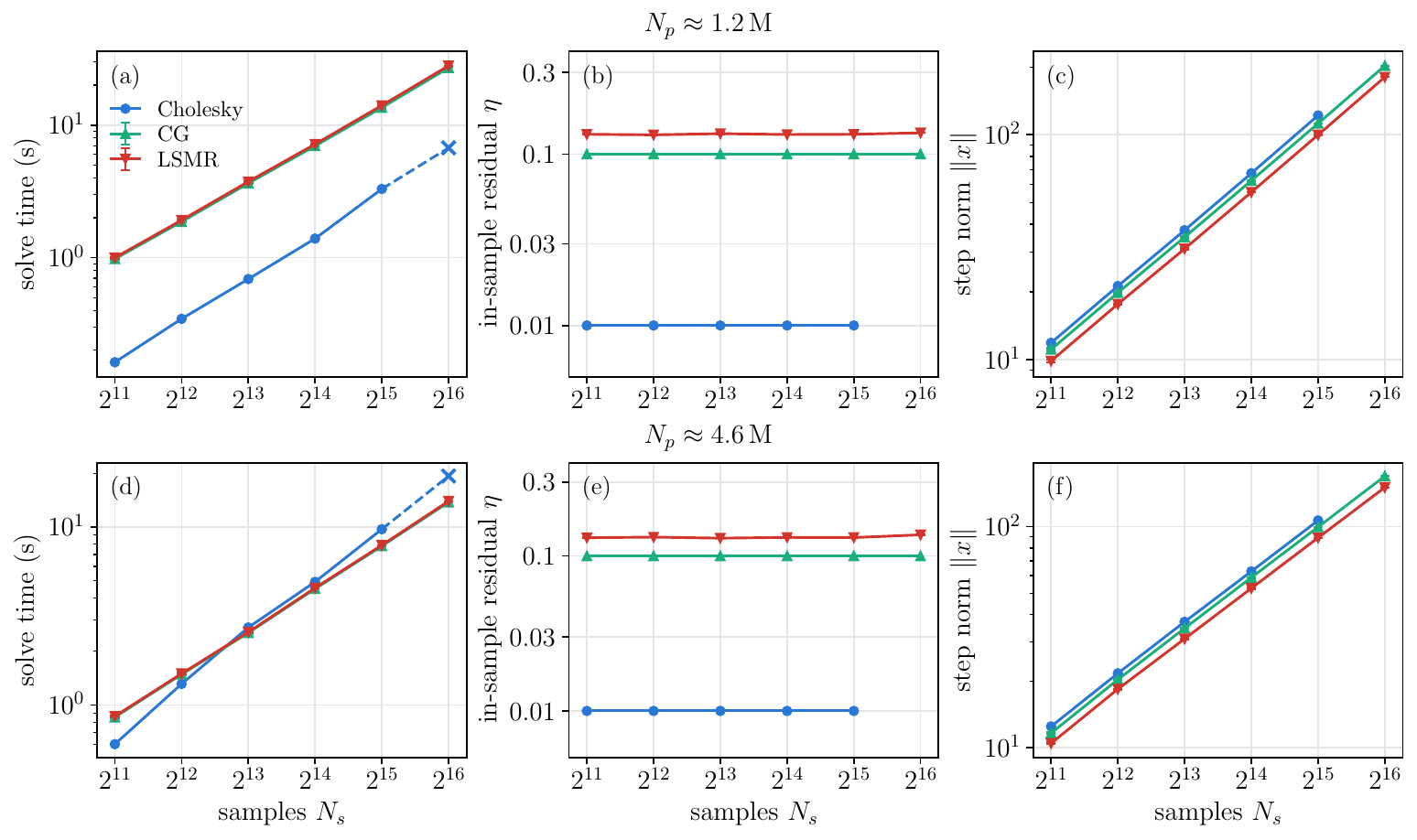}
    \caption{
        Linear solvers on the $8 \times 8$ Hubbard model at $U = 8$, $t' = -0.2$, and $1/8$ hole doping, for (a--c) Pfaffian backflow with $N_p \approx 1.2$M and (d--f) NN-fPEPS with $N_p \approx 4.6$M, using 512 samples per A100 GPU.
        Cholesky runs in FP64, and CG and LSMR in FP32.
        (a),~(d) Time per solve. The cross marks where the FP64 kernel exceeds the GPU memory.
        (b),~(e) In-sample residual, fixed at $\sim 0.1$ for CG, with LSMR run for the same number of iterations as CG, and at $0.01$ for Cholesky by tuning $\lambda$.
        (c),~(f) Step norm.
    }
    \label{fig:sm_solver}
\end{figure}

Fig.\,\ref{fig:sm_solver} compares the three solvers in wall-clock time, in-sample residual $\eta = \lVert A x - b \rVert / \lVert b \rVert$, and step norm $\lVert x \rVert$.
For the Pfaffian backflow with $N_p \approx 1.2$M, the Cholesky solver is four to six times faster than the Krylov solvers at every $N_s$ where its kernel fits in memory, even though it reaches a ten times smaller residual.
This is because the spectrum of $A A^\dagger$ decays quickly for the Pfaffian backflow state, so the Krylov solvers need more iterations (six to eight times more iterations than for the NN-fPEPS).
For the NN-fPEPS with the larger $N_p \approx 4.6$M, the Cholesky solver is faster up to $N_s = 4096$, and the Krylov solvers take over from $N_s \approx 8 \times 10^3$ on.
At the same number of iterations, LSMR leaves a residual $1.3$ times that of CG but returns a step norm $1.1$ times smaller, indicating less overfitting to the finite sample, at nearly the same time cost.

These results suggest choosing the solver by the network size and the degree of parallelism.
The Cholesky solver is the better choice for networks with up to a few million parameters trained on a small number of GPUs, where the $N_s \times N_s$ kernel fits in device memory and its factorization remains cheap.
CG and LSMR are better suited to larger networks distributed over many GPUs, where $N_s$ grows with the number of devices and the factorization becomes prohibitive in both cost and memory, while each Krylov iteration only requires matrix--vector products that parallelize across devices with modest communication.
In this work, we therefore use LSMR for the large-scale parallel simulations of the NN-fPEPS and the Cholesky solver in all other cases.

\subsection{Hyperparameters and details}
\label{sec:sm_hyperparam}

For completeness, we collect here the settings shared by most production runs of this work.

\begin{itemize}
    \item \emph{Sampling.} We employ a mixed sampler that proposes an electron hopping or a neighbor-site exchange with equal probability. The total number of samples is $N_s = 8192$.
    Each VMC iteration includes $2 N$ MCMC steps for system sizes up to $24 \times 24$, and $4 N$ MCMC steps for the $32 \times 32$ lattice.

    \item \emph{Energy clipping.} For robustness against heavy-tailed fluctuations, the local energies entering $\bar{\epsilon}$ are clipped to $\left< E_{\mathrm{loc}} \right> \pm 5\sigma$, with $\sigma^2$ the sampled energy variance~\cite{Gu_NC26_NQS-Hubbard}, while the reported energy and variance are unclipped.

    \item \emph{Optimizer.} In AdamSR, $\mu=0.95$ and $\beta=0.995$. The NN-fPEPS runs adopt the clipped step size of Eq.\,\eqref{eq:sm_clip} with $C_0 = 200$ and the learning rate $\gamma = \max(2.56 / N, 10^{-3})$, while the backflow runs adopt the scheduled step norm of Eq.\,\eqref{eq:sm_constraint} with $C_0 = 1$, reduced to $C_0 = 0.5$ in most runs continued from trained weights, $t_0=1000$, and $T=4000$.

    \item \emph{Linear solver.} When the Cholesky solver is used, the diagonal shift is $\lambda = 10^{-6}$ for the NN-fPEPS and $\lambda = 10^{-3}$ for the backflows. When LSMR is used, the relative tolerance passed to \textsc{lineax} is $10^{-4}$.

    \item \emph{Numerical precision.} The Cholesky solver uses FP64, and the HOTRG steps and the LSMR solver use FP32 without TF32. Other computations are performed in FP32 with TF32 employed whenever possible.

    \item \emph{Weight transfer.} The NN-fPEPS is first trained on the $16 \times 4$ lattice ($t'=0$) or the $8 \times 8$ lattice ($t'=-0.2$). For larger systems, the trained weights are transferred from smaller systems in the following order---$16\times4 \,/\, 8\times8 \rightarrow 16 \times 8 \rightarrow 16 \times 16 \rightarrow 32 \times 32$. Backflows are cold-started at $16 \times 16$, and then transferred to $24 \times 24$ and $32 \times 32$.

    \item \emph{Total iterations.} On the $16 \times 16$ lattice, each state is trained for $\sim 5000$ steps with translation symmetry but without the symmetry projection of Eq.\,\eqref{eq:sm_pointavg}, followed by another $\sim 1000$--$5000$ steps with the projection. On the $32 \times 32$ lattice, this is reduced to $\sim 1000$ steps without projection and $\sim 300$--$1000$ steps with projection.
\end{itemize}

\section{Time cost}

All times are measured on NVIDIA A100-SXM4-80GB GPUs, at batch sizes that saturate the GPU, on the Perlmutter supercomputer at the National Energy Research Scientific Computing Center (NERSC) with 4 GPUs per node, except for the determinant and Pfaffian rows of Table~\ref{tab:sm_amp}, which are taken from the benchmarks of the \textsc{fermix} library~\cite{Chen_github26_fermix} on the same GPU model.
The efficiency is quoted as the model FLOPs utilization
\begin{equation}
    \mathrm{MFU} = \frac{F}{t \, P} \,,
    \label{eq:sm_mfu}
\end{equation}
with $F$ the number of floating-point operations executed by the timed implementation, $t$ the wall-clock time, and $P$ the A100 peak throughput, $156$, $19.5$, and $52$~TFLOP/s for TF32, FP32, and 3$\times$TF32, the emulation of FP32 matrix products by three TF32 products.
For rows that mix several arithmetic types, $t \, P$ is replaced by the sum of $t_i P_i$ over the parts.

\subsection{Amplitude evaluation}
\label{sec:sm_time_amplitude}

\begin{figure}[b]
    \includegraphics[width=0.5\textwidth]{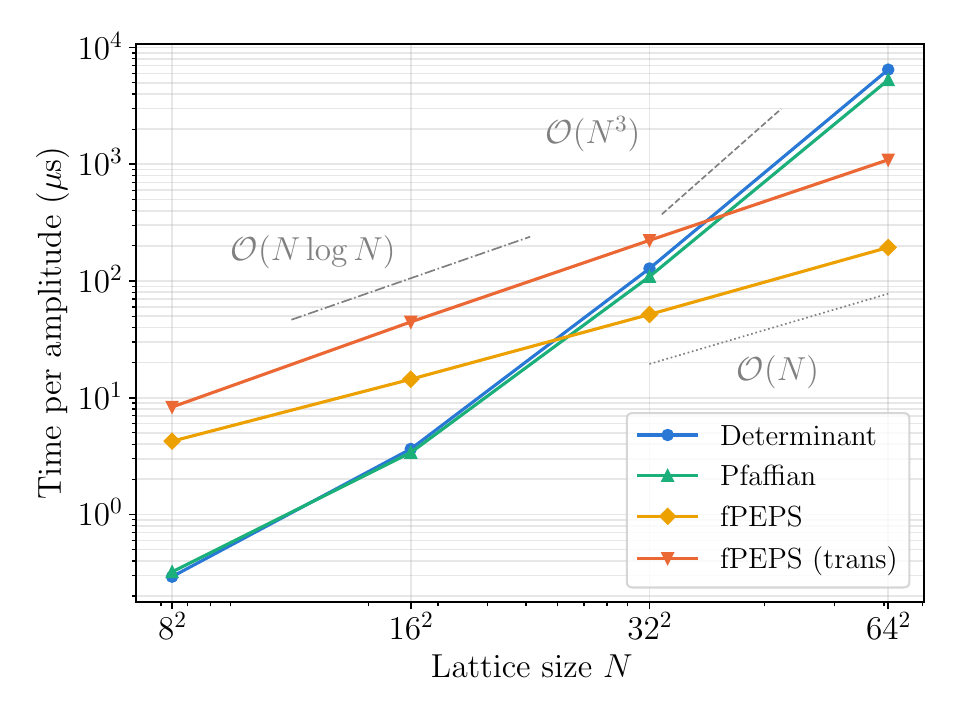}
    \caption{
        Wall-clock time per amplitude evaluation as a function of the lattice size $N = L \times L$, for a Slater determinant, a Pfaffian, an fPEPS with $D = \chi = 4$ at half filling, and the translation-restored fPEPS (trans) of Sec.\,\ref{sec:sm_hotrg_trans}, taken from Table~\ref{tab:sm_amp}.
        Gray guide lines beside the curves indicate the $\mathcal{O}(N^3)$ scaling of the determinant and the Pfaffian, the $\mathcal{O}(N)$ scaling of the fPEPS, and the $\mathcal{O}(N \log N)$ scaling of the fPEPS (trans).
    }
    \label{fig:sm_scaling}
\end{figure}

\begin{table}[p]
    \caption{
        Cost of one amplitude evaluation on one A100-80GB GPU, for the four lattice sizes, with the wall-clock time $t$, the floating-point operation count $F$, the achieved throughput $F / t$, the arithmetic type, and the MFU of Eq.\,\eqref{eq:sm_mfu}.
        The fPEPS has $D = \chi = 4$, and the fPEPS (trans) is the translation-restored contraction of Sec.\,\ref{sec:sm_hotrg_trans}.
        The gray rows split an fPEPS evaluation into the HOTRG steps of Sec.\,\ref{sec:sm_step} and the final exact contraction of the remaining $L \times 4$ torus in Eq.\,\eqref{eq:sm_trace}, and the bottleneck and normal CNNs are the networks of Fig.\,\ref{fig:sm_network}.
    }
    \label{tab:sm_amp}
    \renewcommand{\arraystretch}{0.96}
    \begin{ruledtabular}
    \small
    \begin{tabular}{lccccc}
        \amphead \\
        \colrule
        \ampsize{$8 \times 8$} \\
        \amprow{Determinant}{$0.293$}{$0.175$}{$0.60$}{3$\times$TF32}{$1.1\%$} \\
        \amprow{Pfaffian}{$0.322$}{$0.0874$}{$0.27$}{3$\times$TF32}{$0.52\%$} \\
        \amprow{fPEPS}{$4.24$}{$58.1$}{$13.7$}{mixed}{$13\%$} \\
        \amprowgr{\subrow{HOTRG steps}}{$1.50$}{$2.75$}{$1.83$}{FP32}{$9.4\%$} \\
        \amprowgr{\subrow{Final exact contraction}}{$2.73$}{$55.3$}{$20.2$}{TF32}{$13\%$} \\
        \amprow{fPEPS (trans)}{$8.32$}{$116$}{$14.0$}{mixed}{$13\%$} \\
        \amprowgr{\subrow{HOTRG steps}}{$2.86$}{$5.49$}{$1.92$}{FP32}{$9.9\%$} \\
        \amprowgr{\subrow{Final exact contraction}}{$5.44$}{$111$}{$20.4$}{TF32}{$13\%$} \\
        \amprow{Bottleneck CNN}{$14.9$}{$579$}{$39.0$}{TF32}{$25\%$} \\
        \amprow{Normal CNN}{$4.28$}{$152$}{$35.6$}{TF32}{$23\%$} \\
        \colrule
        \ampsize{$16 \times 16$} \\
        \amprow{Determinant}{$3.64$}{$11.2$}{$3.07$}{3$\times$TF32}{$5.9\%$} \\
        \amprow{Pfaffian}{$3.39$}{$5.59$}{$1.65$}{3$\times$TF32}{$3.2\%$} \\
        \amprow{fPEPS}{$14.4$}{$144$}{$10.0$}{mixed}{$14\%$} \\
        \amprowgr{\subrow{HOTRG steps}}{$8.74$}{$16.5$}{$1.89$}{FP32}{$9.7\%$} \\
        \amprowgr{\subrow{Final exact contraction}}{$5.62$}{$127$}{$22.7$}{TF32}{$15\%$} \\
        \amprow{fPEPS (trans)}{$44.6$}{$554$}{$12.4$}{mixed}{$14\%$} \\
        \amprowgr{\subrow{HOTRG steps}}{$22.6$}{$43.9$}{$1.95$}{FP32}{$10\%$} \\
        \amprowgr{\subrow{Final exact contraction}}{$22.1$}{$510$}{$23.0$}{TF32}{$15\%$} \\
        \amprow{Bottleneck CNN}{$57.1$}{$2316$}{$40.6$}{TF32}{$26\%$} \\
        \amprow{Normal CNN}{$16.0$}{$609$}{$38.1$}{TF32}{$24\%$} \\
        \colrule
        \ampsize{$32 \times 32$} \\
        \amprow{Determinant}{$128$}{$716$}{$5.61$}{3$\times$TF32}{$11\%$} \\
        \amprow{Pfaffian}{$109$}{$358$}{$3.27$}{3$\times$TF32}{$6.3\%$} \\
        \amprow{fPEPS}{$51.7$}{$349$}{$6.75$}{mixed}{$14\%$} \\
        \amprowgr{\subrow{HOTRG steps}}{$40.3$}{$76.9$}{$1.91$}{FP32}{$9.8\%$} \\
        \amprowgr{\subrow{Final exact contraction}}{$11.4$}{$272$}{$23.8$}{TF32}{$15\%$} \\
        \amprow{fPEPS (trans)}{$223$}{$2438$}{$10.9$}{mixed}{$15\%$} \\
        \amprowgr{\subrow{HOTRG steps}}{$134$}{$264$}{$1.96$}{FP32}{$10\%$} \\
        \amprowgr{\subrow{Final exact contraction}}{$89.9$}{$2174$}{$24.2$}{TF32}{$15\%$} \\
        \amprow{Bottleneck CNN}{$219$}{$9263$}{$42.3$}{TF32}{$27\%$} \\
        \amprow{Normal CNN}{$52.4$}{$2437$}{$46.5$}{TF32}{$30\%$} \\
        \colrule
        \ampsize{$64 \times 64$} \\
        \amprow{Determinant}{$6469$}{$4.58 \times 10^{4}$}{$7.08$}{3$\times$TF32}{$14\%$} \\
        \amprow{Pfaffian}{$5281$}{$2.29 \times 10^{4}$}{$4.34$}{3$\times$TF32}{$8.3\%$} \\
        \amprow{fPEPS}{$194$}{$892$}{$4.60$}{mixed}{$13\%$} \\
        \amprowgr{\subrow{HOTRG steps}}{$171$}{$330$}{$1.93$}{FP32}{$9.9\%$} \\
        \amprowgr{\subrow{Final exact contraction}}{$23.4$}{$560$}{$23.9$}{TF32}{$15\%$} \\
        \amprow{fPEPS (trans)}{$1093$}{$1.04 \times 10^{4}$}{$9.49$}{mixed}{$15\%$} \\
        \amprowgr{\subrow{HOTRG steps}}{$735$}{$1407$}{$1.91$}{FP32}{$9.8\%$} \\
        \amprowgr{\subrow{Final exact contraction}}{$365$}{$8964$}{$24.6$}{TF32}{$16\%$} \\
        \amprow{Bottleneck CNN}{$868$}{$3.71 \times 10^{4}$}{$42.7$}{TF32}{$27\%$} \\
        \amprow{Normal CNN}{$206$}{$9749$}{$47.4$}{TF32}{$30\%$} \\
    \end{tabular}
    \end{ruledtabular}
\end{table}

Table~\ref{tab:sm_amp} and Fig.\,\ref{fig:sm_scaling} give the cost of one evaluation of each component of the ansatzes on a single GPU.
An NN-fPEPS amplitude combines the bottleneck CNN with the fPEPS (trans), with comparable shares at every size.
A Pfaffian backflow amplitude combines the normal CNN with the Pfaffian, apart from the untimed assembly of the pairing matrix, and is dominated by the normal CNN up to $16 \times 16$ and by the Pfaffian from $32 \times 32$ on.
The determinant and Pfaffian rows use our batched GPU kernels for the blocked LU and Parlett--Reid factorizations of the \textsc{fermix} library~\cite{Chen_github26_fermix}.
All rows are measured at half filling.

\subsection{VMC step}
\label{sec:sm_time_step}

\begin{figure}[tp]
    \includegraphics[width=\textwidth]{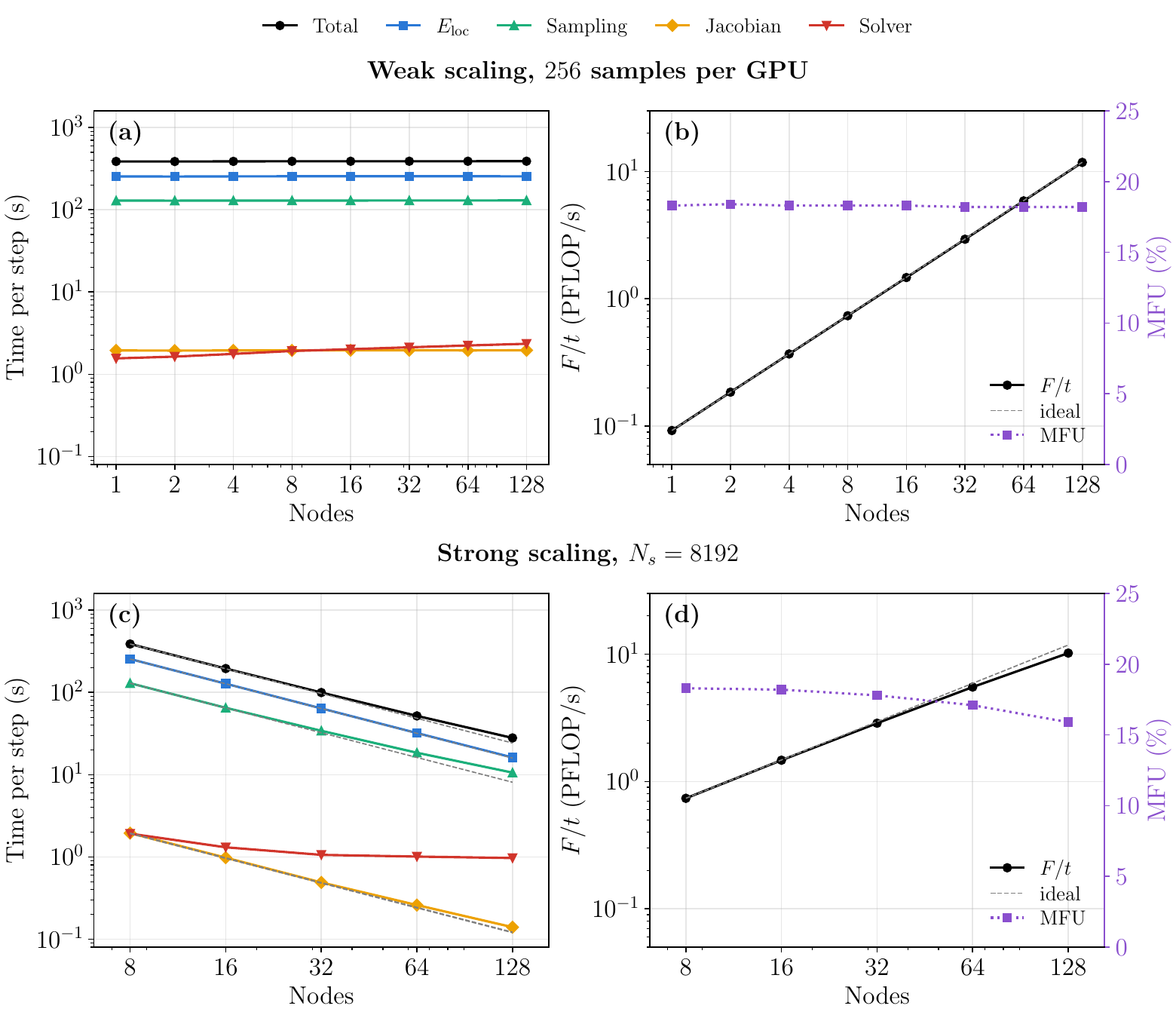}
    \caption{
        Weak (a, b) and strong (c, d) scaling of one VMC step of the NN-fPEPS on the $16 \times 16$ lattice.
        (a, c) Wall-clock time of the four parts of a step and of the whole step.
        (b, d) Aggregate throughput $F / t$ over all GPUs, with the dashed line its ideal linear scaling through the first point, and the MFU of Eq.\,\eqref{eq:sm_mfu} on the right axis.
    }
    \label{fig:sm_vmc_scaling}
\end{figure}

Fig.\,\ref{fig:sm_vmc_scaling} shows the weak and strong scaling of one optimization step of the $16 \times 16$ NN-fPEPS production runs at $U = 8$ and $t' = -0.2$.
A step consists of one sweep of $2N$ Metropolis updates per sample, the local energies with about $2N$ amplitudes $\psi(n')$ per sample, and the Jacobian at twice the cost of a forward pass, all evaluated with the $8$ elements of the symmetry group $P$ in one call.
The number of LSMR iterations is fixed at $100$ for a direct comparison, although it may need to grow with $N_s$ to keep the residual error $\eta$ unchanged.
The tests stop at $512$ GPUs due to the limited number of A100-80GB nodes on Perlmutter. The $1024$-GPU runs in this work use A100-40GB nodes.

\section{Pure fPEPS reference}
\label{sec:sm_su}

We estimate the exact ground-state energy of the $16 \times 16$ PBC lattice at $t' = 0$ with pure fPEPS optimized by simple update (SU)~\cite{PhysRevLett.101.090603,PhysRevLett.98.070201}, evaluated by contracting finite clusters embedded in a belief-propagation environment, and extrapolated to infinite bond dimension.

\subsection{Simple update and cluster contraction}

\begin{figure}
    \centering
    \includegraphics[width=\linewidth]{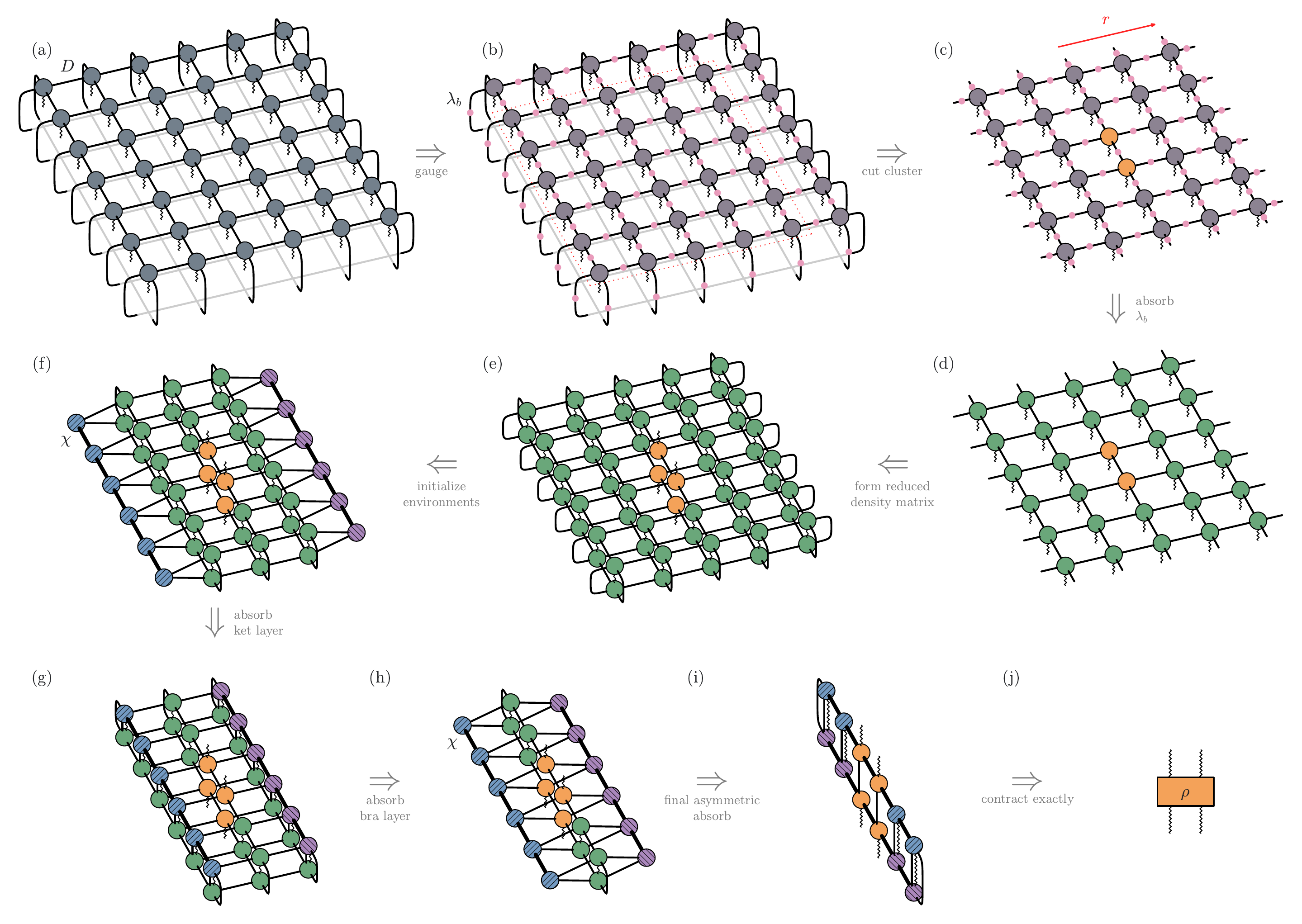}
    \caption{
    Periodic PEPS calculations with the `cluster boundary' method.
    (a) An initial periodic PEPS with bond dimension $D$, physical legs denoted by zigzags.
    (b) After SU gauging, each bond carries a bond weight $\lambda_b$.
    (c) A rectangular cluster of tensors is cut out up to $r$ sites in each direction from the target sites (orange), here $r=2$. Every bond, including those on the boundary, gets a bond weight.
    (d) The bond weights are absorbed to yield a single tensor per site.
    (e) We form the partial trace network, with the environment beyond the cluster now equivalent to the belief propagation approximation.
    (f) We form left and right environments (hatched blue and purple tensors) parallel to the target sites, truncating to bond dimension $\chi$.
    (g) Each environment gets compressed with a single inward layer of ket tensors.
    (h) Each environment then gets compressed with the corresponding bra layer of tensors.
    For larger $r$, steps (f)--(h) are repeated until each environment is adjacent to the target sites.
    (i) Finally, the left environment is compressed with the ket layer of the target column, and the right environment is compressed with the bra layer.
    (j) This gives a ladder-like tensor network, which can be contracted exactly with $\mathcal{O}(D^3 \chi^3)$ cost to yield the reduced density matrix $\rho$ of the target sites.
    Steps (c)--(j) are repeated for each target reduced density matrix.
    }
    \label{fig:fpeps-cluster-diagram}
\end{figure}

\label{sec:sm_su_contraction}
The fPEPS use the local fermionic encoding~\cite{gu2010grassmann,Mortier_SP25_fTN,Gao_PRR25_fTN-contraction} with $U(1)$ symmetry~\cite{PhysRevA.82.050301,PhysRevB.83.115125,PhysRevB.83.125106}.
The initial product state is a period-$8$ stripe, with antiferromagnetic order on columns $1$--$7$ and $9$--$15$, a $\pi$ phase shift between the two domains, and empty columns $8$ and $16$ that hold the $32$ holes of $1/8$ doping.
We first converge SU at bond dimension $D = 6$ and then continue from this state directly at each target bond dimension $D = 8, 10, 12, 14, 16$.
All runs use a second-order Trotter decomposition with the imaginary time step $\tau = 0.5 D^{-3/2}$.
All calculations are performed on CPUs with \textsc{quimb}~\cite{Gray_JOSS18_quimb} and the \textsc{symmray} array backend~\cite{Gray_github25_symmray}.

To evaluate observables of the SU states on the torus, we first gauge each state by running SU without applying any gates, which is equivalent to belief propagation (BP)~\cite{PhysRevResearch.3.023073,SciPostPhys.15.6.222}.
This yields a bond weight $\lambda_b$ on each of the $N_b = 2L^2 = 512$ bonds, shown in Fig.\,\ref{fig:fpeps-cluster-diagram}(a)--(b).

For each nearest-neighbor term, acting on sites $(i, j)$ and $(i, j+1)$ or $(i, j)$ and $(i+1, j)$, we build a rectangular cluster that extends $r$ sites in every direction from the two target sites.
A horizontal term thus has a $C_x \times C_y = 4 \times 3$ cluster at $r = 1$ and a $16 \times 15$ cluster, which covers nearly the whole lattice, at the largest radius $r = 7$.
We cut the bonds along the cluster boundary and place the weights $\lambda_b$ on every bond of the cluster, including the dangling bonds at its edges.
When the cluster spans the full lattice in one direction, $C_x = 16$ or $C_y = 16$, its two far edges meet, and both ends of the cut bond between them carry $\lambda_b$.
This amounts to approximating the environment beyond the cluster by its BP form.
This process is shown in Fig.\,\ref{fig:fpeps-cluster-diagram}(b)--(d).

Each cluster is a PEPS with open boundary conditions, which we partially trace by boundary contraction to obtain an approximate reduced density matrix of the two target sites, as shown in Fig.\,\ref{fig:fpeps-cluster-diagram}(e).
The two boundary environments start from the cluster edges parallel to the target bond and sweep inwards, with the ket and bra layers absorbed in two separate steps and the environments compressed to bond dimension $\chi$ after each step.
Once the two environments are adjacent to the target site column, one environment is compressed with the ket tensors, and the other with the bra tensors.
See Fig.\,\ref{fig:fpeps-cluster-diagram}(f)--(i).
The compression is a variant of successive deterministic compression~\cite{milbradt2026efficient}.
A sketching sweep with randomized low-rank SVDs first compresses the environment to the oversampled bond dimension $1.5 \chi$, truncating every charge sector in proportion, and a final direct SVD then truncates it to $\chi$ by ranking the singular values across all charge sectors.
The contraction thus costs $\mathcal{O}(D^3 \chi^3)$, remains accurate, and is compatible with $U(1)$ fermionic tensor networks.
Each reduced density matrix is normalized individually.

\begin{figure}[h]
    \centering
    \includegraphics[width=\linewidth]{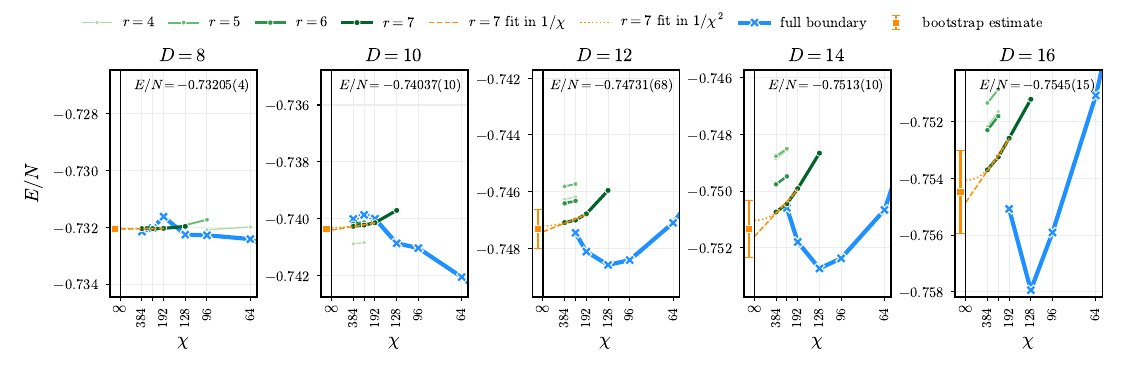}
    \caption{Convergence of the pure fPEPS energy in bond dimension $D$. Green lines with circles: energy calculations with respect to boundary bond dimension $\chi$ and cluster radius $r$.
    Blue line with crosses: energy calculated with full periodic boundary contraction.
    Orange lines: $r=7$ linear and quadratic fits of cluster boundary estimates. Orange marker: bootstrap estimated energy per $D$, including error accounting for uncertainty of fit and convergence in $r$.
    Error bars show one standard deviation.
    }
    \label{fig:fpeps-convergence}
\end{figure}

The results for the energy per site, $E/N$, as a function of $D$, $r$, and $\chi$ are shown in Fig.\,\ref{fig:fpeps-convergence}. We also show the energies calculated by full boundary contraction, i.e., without the cluster approximation. These are more expensive to compute, with a cost of $\mathcal{O}(D^5 \chi^3)$, and converge more noisily, so we do not use them to estimate energies. Nonetheless, they appear to be consistent with the cluster boundary approximation.

\subsection{Extrapolation}
\label{sec:sm_su_extrap}

The procedure above gives the energy per site as a function of the PEPS bond dimension $D$, the cluster radius $r$, and the boundary bond dimension $\chi$, which we extrapolate by bootstrapping.

\begin{figure}[h]
    \centering
    \includegraphics[width=0.5\linewidth]{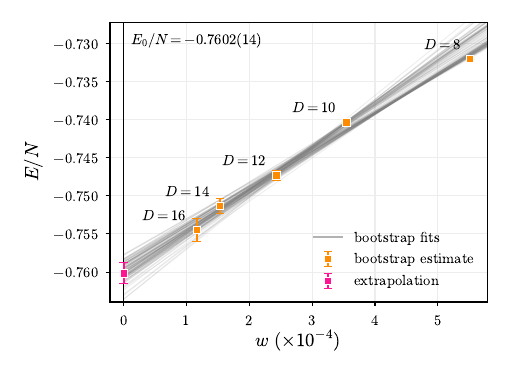}
    \caption{Bootstrap extrapolation of the $D=\infty$ energy $E_0$ in the gauge weight $w$.
    Orange markers show individual bootstrap energy estimates for each $D$.
    Gray lines show a sample of 100 fitted bootstrap lines.
    The pink marker shows the mean value of all $10{,}000$ extrapolated bootstrap samples, with standard deviation.
    }
    \label{fig:fpeps-w-extrap}
\end{figure}

For each $D$, we fit the energies at the largest radius $r = 7$ and the three largest boundary dimensions $\chi = 192$, $256$, and $384$ to
\begin{equation}
    E(\chi) = E_\infty + a \chi^{-p} \,,
    \label{eq:sm_su_chi_fit}
\end{equation}
with the exponent $p \in \{1, 2\}$ drawn at random.
The cluster approximation error is accounted for by adding Gaussian noise to $E_\infty$ with the standard deviation
\begin{equation}
    \sigma_r = \left| E(r = 7, \chi = 384) - E(r = 6, \chi = 384) \right| \,.
    \label{eq:sm_su_cluster_noise}
\end{equation}
Individual estimates for $E/N$ at each $D$ are shown in Figs.\,\ref{fig:fpeps-convergence} and \ref{fig:fpeps-w-extrap}. Note that these do not enter the overall bootstrap procedure.

To extrapolate in $D$, we use a proxy for the SU discarded weight.
The gauge weight of bond $b$ is the smallest normalized squared singular value
\begin{equation}
    w_b = \frac{\min_k \lambda_b^2[k]}{\sum_k \lambda_b^2[k]} \,,
    \label{eq:sm_su_gauge_weight}
\end{equation}
where $k$ runs over the singular values in all charge sectors, and the gauge weight of a state is the geometric mean over all bonds $w = \left( \prod_b w_b \right)^{1/N_b}$.
We then extract the energy $E_0$ from the linear fit
\begin{equation}
    E(w) = E_0 + c\, w \,.
    \label{eq:sm_su_w_fit}
\end{equation}
Note that the quantity $w$ and the fitting form here differ slightly from those in, e.g., Ref.~\cite{PhysRevB.93.045116}.

A single bootstrap sample draws $p$ and the cluster noise for each $D = 8, 10, 12, 14, 16$, keeps or drops the $D = 8$ point with equal probability, and fits Eq.\,\eqref{eq:sm_su_w_fit} to the remaining points.
The mean and standard deviation over $10{,}000$ bootstrap samples give $E/N = -0.7602(14)$.
Fig.\,\ref{fig:fpeps-w-extrap} shows this estimate with 100 fitted line samples, and the separate individual estimates for each $D$, as a function of $w$.

\clearpage

\section{Energy benchmarks}
\label{sec:sm_energy}

\subsection{Zero-variance extrapolation}
\label{sec:sm_zve}
Close to an eigenstate, the energy of a variational state approaches the exact energy linearly in its energy variance, $E \approx E_0 + k \sigma^2$~\cite{Sorella_PRB01_Lanczos,Becca_17_VMCtext}.
The states along a VMC optimization therefore fall on a straight line in the $(\sigma^2, E)$ plane, whose intercept at $\sigma^2 = 0$ estimates the exact ground-state energy $E_0$.

\begin{figure}[t]
    \centering
    \includegraphics[width=\textwidth]{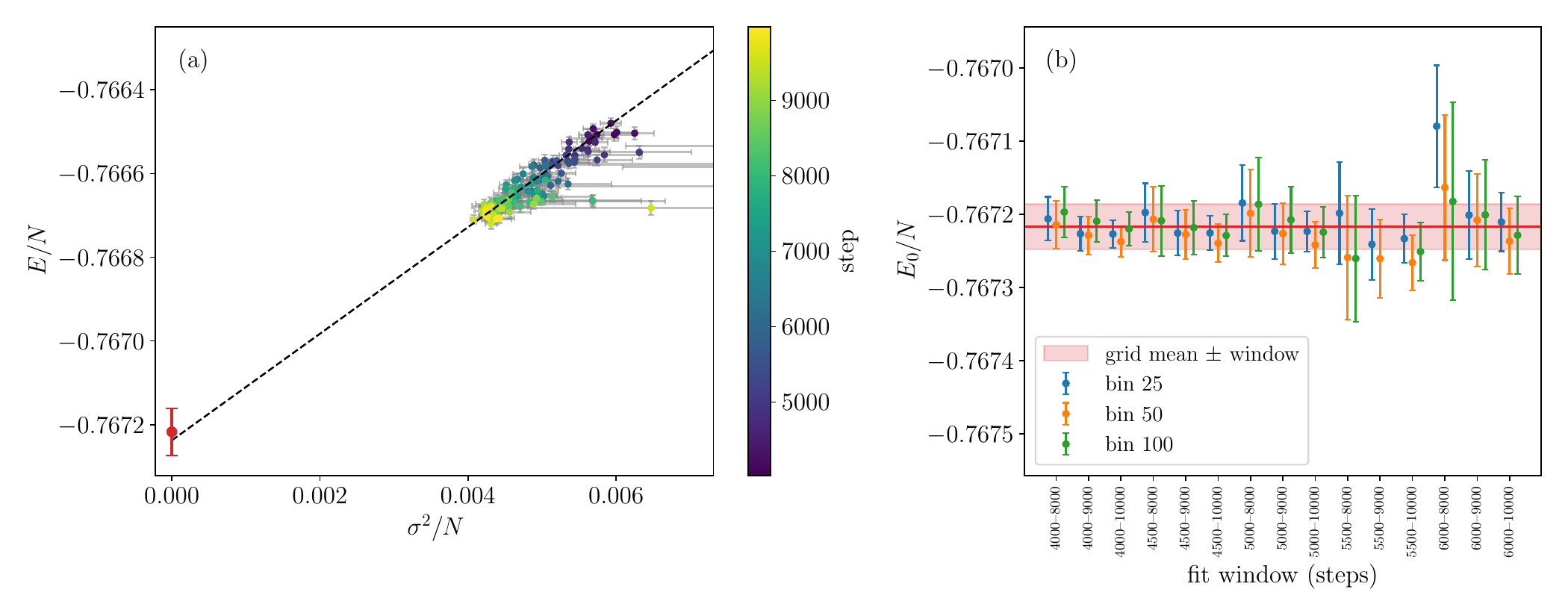}
    \caption{Zero-variance extrapolation on the $16 \times 4$ lattice at $t' = 0$.
    (a) Energy against variance of the training curve in bins of $50$ steps between steps $4000$ and $10{,}000$, colored by the step, with the fitted line (dashed) and the extrapolated $E_0$ of (b) (red). A few bins with variance spikes lie outside the plotted range but remain in the fit.
    (b) Intercepts of the fits with the windows (horizontal axis) and bin sizes (colors).
    }
    \label{fig:sm_zve}
\end{figure}

This technique is used to estimate an accurate ground-state energy on the $16 \times 4$ lattice at $t' = 0$ using an NN-fPEPS with $N_p \approx 38$M, much larger than that of Table~\ref{tab:sm_network}.
The symmetry projection of Eq.\,\eqref{eq:sm_pointavg} is also used during training for better accuracy.
This state serves only for the extrapolation and the variational energy is not listed in Table~\ref{tab:sm_energy}.
We bin the energy and variance of its training curve and fit a straight line by orthogonal distance regression, which accounts for the statistical errors of both (Fig.\,\ref{fig:sm_zve}(a)).
Repeating the fit over a grid of step windows and bin sizes (Fig.\,\ref{fig:sm_zve}(b)), we obtain $E_0/N = -0.76722(6)$.

\subsection{Variational energy}
Table~\ref{tab:sm_energy} lists the variational energies and energy variances of all states in this work, before and after the symmetry projection, together with selected variational energies in existing literature.
All our states except the pure fPEPS are translation invariant, and each projected state continues the optimization of the unprojected one under Eq.\,\eqref{eq:sm_pointavg}.
The projected energies in Figs.\,3(a) and 4(a) of the main text are taken from this table.

\begin{table}[b]
    \caption{
        Variational energy per site $E/N$ and energy variance per site $\sigma^2/N = (\langle H^2 \rangle - E^2)/N$ of the Hubbard model at $U = 8$ and $1/8$ doping on $L_x \times L_y$ lattices with PBC, without and with the symmetry projection.
        Rows with a reference are taken from the literature, including the hidden fermion Pfaffian state (HFPS) and variational auxiliary-field quantum Monte Carlo (VAFQMC), and all other rows are our own states.
        Our projection is onto the mirror and spin-inversion group $P$ of Eq.\,\eqref{eq:sm_symm_pointgroup}, and that of Ref.~\cite{Rende_arxiv26_SBP} onto the point group $C_{4v}$.
        On the $16 \times 4$ lattice, HFPS, ACE, and the neural network backflow (NNBF) are projected onto the full mirror, translation, and spin-inversion symmetry.
        The pure fPEPS rows are the simple-update states of Sec.\,\ref{sec:sm_su}.
        The rows marked red are extrapolations that estimate the exact energy, the zero-variance extrapolation of Sec.\,\ref{sec:sm_zve} on $16 \times 4$ and the pure fPEPS extrapolation of Sec.\,\ref{sec:sm_su_extrap} on $16 \times 16$.
        Values marked with $\dagger$ are averages over the last $50$ to $100$ optimization steps instead of measurements.
        For each lattice and $t'$, the lowest variational energy and the lowest variance are shown in bold.
    }
    \label{tab:sm_energy}
    \newcommand{\best}[1]{\textbf{\boldmath$#1$}}
    \newcommand{\extrap}[1]{\textcolor{red}{#1}}
    \begin{ruledtabular}
    \begin{tabular}{ll|cccc}
        & & \multicolumn{2}{c}{Without projection} & \multicolumn{2}{c}{With projection} \\
        \cline{3-4} \cline{5-6}
        Lattice & Ansatz & $E/N$ & $\sigma^2/N$ & $E/N$ & $\sigma^2/N$ \\
        \colrule\noalign{\vskip\doublerulesep}\hline
        \multicolumn{6}{c}{$t' = 0$} \\
        \colrule
        $16 \times 4$ & Transformer det backflow~\cite{Gu_NC26_NQS-Hubbard} & $-0.76298$ & --- & --- & --- \\
        & HFPS~\cite{Chen_PNAS26_HFPS} & --- & --- & $-0.76413(3)$ & $0.0313$ \\
        & ACE (det backflow)~\cite{Gu_PRX-Intelligence_SCALE} & $-0.76464(1)$ & --- & $-0.76623(1)$ & --- \\
        & NNBF (det backflow)~\cite{Loehr_arxiv25_NNBF} & --- & --- & $-0.766073(6)$ & --- \\
        & NN-fPEPS & $-0.76481^\dagger$ & $0.017^\dagger$ & \best{-0.766455(14)} & \best{0.00555(9)} \\
        & \extrap{NN-fPEPS (extrapolated)} & \extrap{---} & \extrap{---} & \extrap{$-0.76722(6)$} & \extrap{---} \\
        \colrule
        $16 \times 16$ & Transformer det backflow~\cite{Gu_NC26_NQS-Hubbard} & $-0.7563$ & --- & --- & --- \\
        & HFPS~\cite{Roth_arxiv25_HubbardSC} & --- & --- & $-0.7515(1)$ & $0.0638$ \\
        & SCALE (det backflow)~\cite{Gu_PRX-Intelligence_SCALE} & $-0.7529$ & --- & --- & --- \\
        & ACE (det backflow)~\cite{Gu_PRX-Intelligence_SCALE} & $-0.7573$ & --- & --- & --- \\
        & VAFQMC~\cite{Sorella_PRB23_VAFQMC, Wu_Science24_Vscore} & $-0.75865(3)$ & \best{0.0059(8)} & --- & --- \\
        & fPEPS ($D = 16$) & $-0.7545(15)$ & --- & --- & --- \\
        & BCS backflow & $-0.75772(1)$ & $0.0308(6)$ & \best{-0.75878(2)} & $0.0255(20)$ \\
        & Pfaffian backflow & $-0.75686(2)$ & $0.0337(5)$ & $-0.75813(1)$ & $0.0259(2)$ \\
        & NN-fPEPS & $-0.75560(4)$ & $0.0379(14)$ & $-0.75851(2)$ & $0.0256(6)$ \\
        & \extrap{fPEPS (extrapolated)} & \extrap{$-0.7602(14)$} & \extrap{---} & \extrap{---} & \extrap{---} \\
        \colrule
        $32 \times 32$ & SCALE (det backflow)~\cite{Gu_PRX-Intelligence_SCALE} & $-0.7501$ & --- & --- & --- \\
        & BCS backflow & $-0.75490^\dagger$ & $0.067^\dagger$ & $-0.75617(1)$ & $0.046(10)$ \\
        & NN-fPEPS & $-0.75511^\dagger$ & $0.054^\dagger$ & \best{-0.75677(1)} & \best{0.0346(4)} \\
        \colrule\noalign{\vskip\doublerulesep}\hline
        \multicolumn{6}{c}{$t' = -0.2$} \\
        \colrule
        $12 \times 12$ & Transformer det backflow~\cite{Gu_NC26_NQS-Hubbard} & $-0.7414$ & --- & --- & --- \\
        & BCS backflow~\cite{Rende_arxiv26_SBP} & --- & --- & \best{-0.74425(1)} & \best{0.035(1)} \\
        \colrule
        $16 \times 16$ & HFPS~\cite{Roth_arxiv25_HubbardSC} & --- & --- & $-0.7360(1)$ & $0.0762$ \\
        & SCALE (det backflow)~\cite{Gu_PRX-Intelligence_SCALE} & $-0.7373$ & --- & --- & --- \\
        & ACE (det backflow)~\cite{Gu_PRX-Intelligence_SCALE} & $-0.7430$ & --- & --- & --- \\
        & BCS backflow~\cite{Rende_arxiv26_SBP} & --- & --- & $-0.74411(1)$ & $0.037(1)$ \\
        & Pfaffian backflow & $-0.74372(1)$ & $0.0346(5)$ & \best{-0.74533(1)} & \best{0.0272(5)} \\
        & NN-fPEPS & $-0.74144(4)$ & $0.0474(8)$ & $-0.74467(2)$ & $0.0367(31)$ \\
        \colrule
        $20 \times 20$ & BCS backflow~\cite{Rende_arxiv26_SBP} & --- & --- & \best{-0.74382(1)} & \best{0.041(1)} \\
        \colrule
        $24 \times 24$ & BCS backflow~\cite{Rende_arxiv26_SBP} & --- & --- & $-0.74397(1)$ & $0.054(1)$ \\
        & Pfaffian backflow & $-0.74457(2)$ & $0.0358(4)$ & \best{-0.74585(1)} & \best{0.0257(3)} \\
        \colrule
        $32 \times 32$ & Pfaffian backflow & $-0.74453^\dagger$ & $0.048^\dagger$ & \best{-0.74569(1)} & \best{0.0294(5)} \\
        & NN-fPEPS & $-0.74032^\dagger$ & $0.081^\dagger$ & $-0.74257(2)$ & $0.0502(7)$ \\
    \end{tabular}
    \end{ruledtabular}
\end{table}

\clearpage
\section{Real-space correlations}
\label{sec:sm_orders}

\begin{figure}[b]
    \centering
    \includegraphics[width=\textwidth]{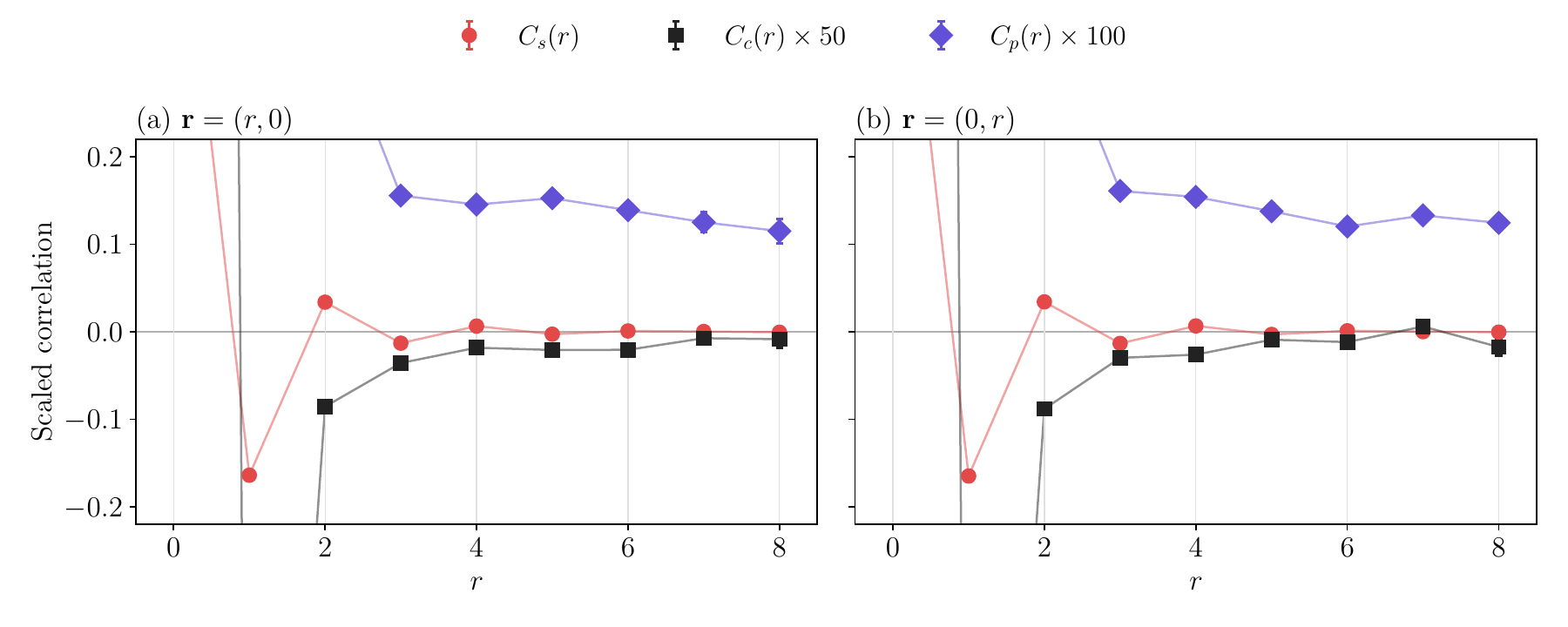}
    \caption{
        Spin correlation $C_s(r)$, charge correlation $C_c(r)$ of Eq.\,\eqref{eq:sm_corr}, and pair correlation $C_p(r)$ of Eq.\,(2) of the main text for the projected BCS backflow state on the $16 \times 16$ lattice at $t' = 0$, along (a) $\bm{r} = (r, 0)$ and (b) $\bm{r} = (0, r)$.
    }
    \label{fig:sm_corr_tp0_16_bcs}
\end{figure}

\begin{figure}[b]
    \centering
    \includegraphics[width=\textwidth]{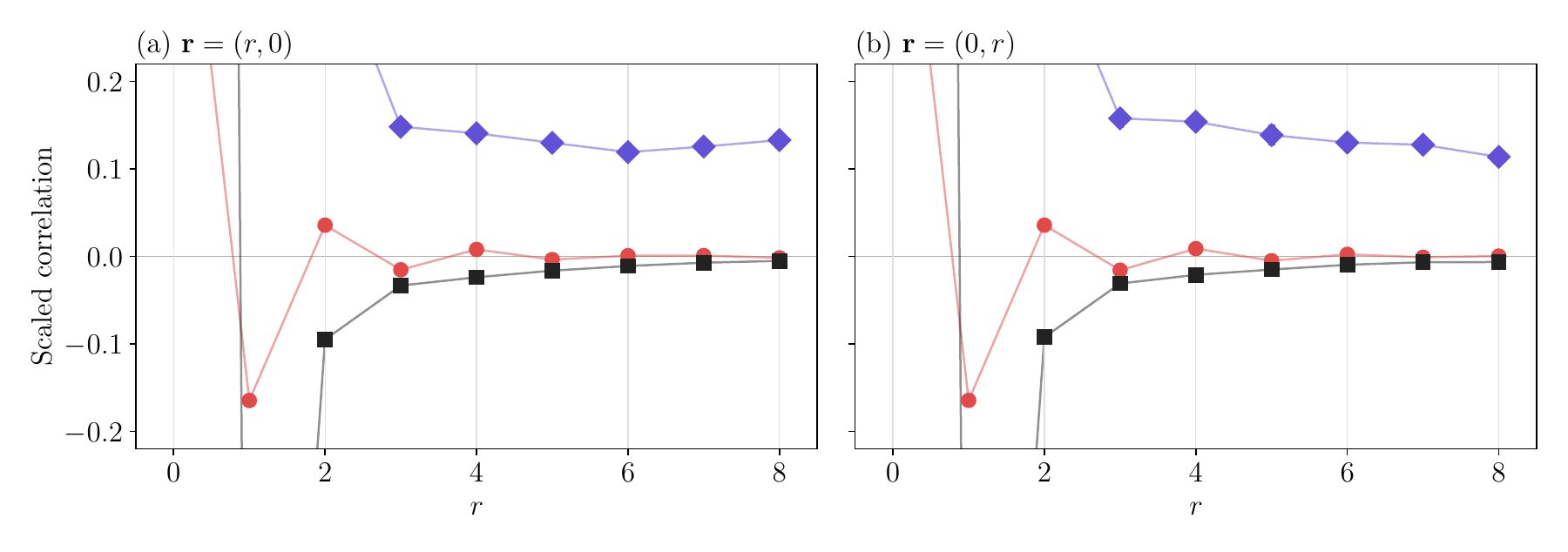}
    \caption{
        Same as Fig.\,\ref{fig:sm_corr_tp0_16_bcs}, for the projected Pfaffian backflow state on the $16 \times 16$ lattice at $t' = 0$.
    }
    \label{fig:sm_corr_tp0_16_pf}
\end{figure}

\begin{figure}[tp]
    \centering
    \includegraphics[width=\textwidth]{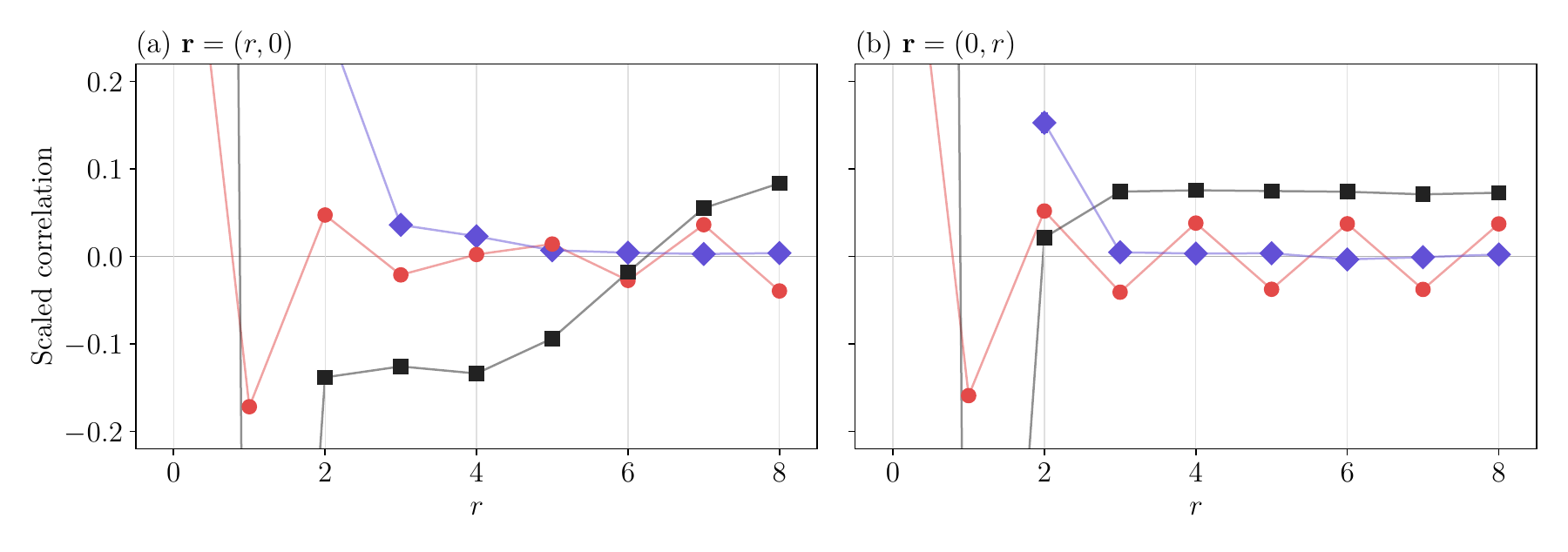}
    \caption{
        Same as Fig.\,\ref{fig:sm_corr_tp0_16_bcs}, for the projected NN-fPEPS on the $16 \times 16$ lattice at $t' = 0$.
    }
    \label{fig:sm_corr_tp0_16_nn}
\end{figure}

\begin{figure}[tp]
    \centering
    \includegraphics[width=\textwidth]{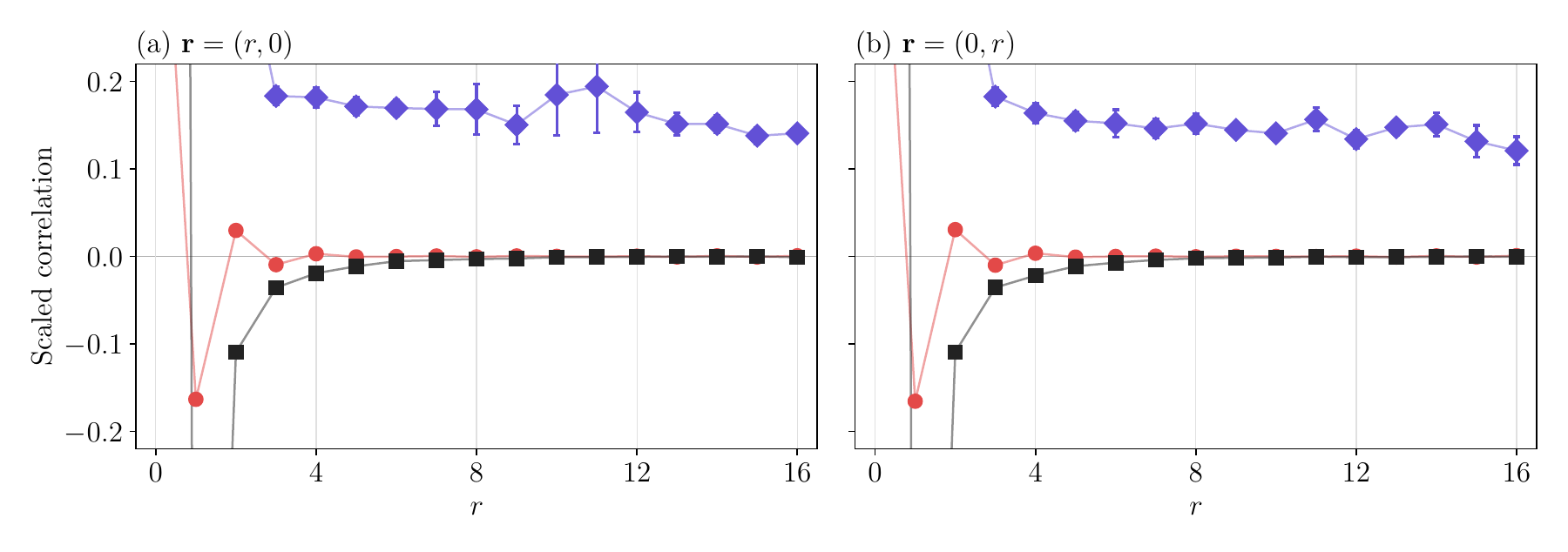}
    \caption{
        Same as Fig.\,\ref{fig:sm_corr_tp0_16_bcs}, for the projected BCS backflow state on the $32 \times 32$ lattice at $t' = 0$.
    }
    \label{fig:sm_corr_tp0_32_bcs}
\end{figure}

\begin{figure}[tp]
    \centering
    \includegraphics[width=\textwidth]{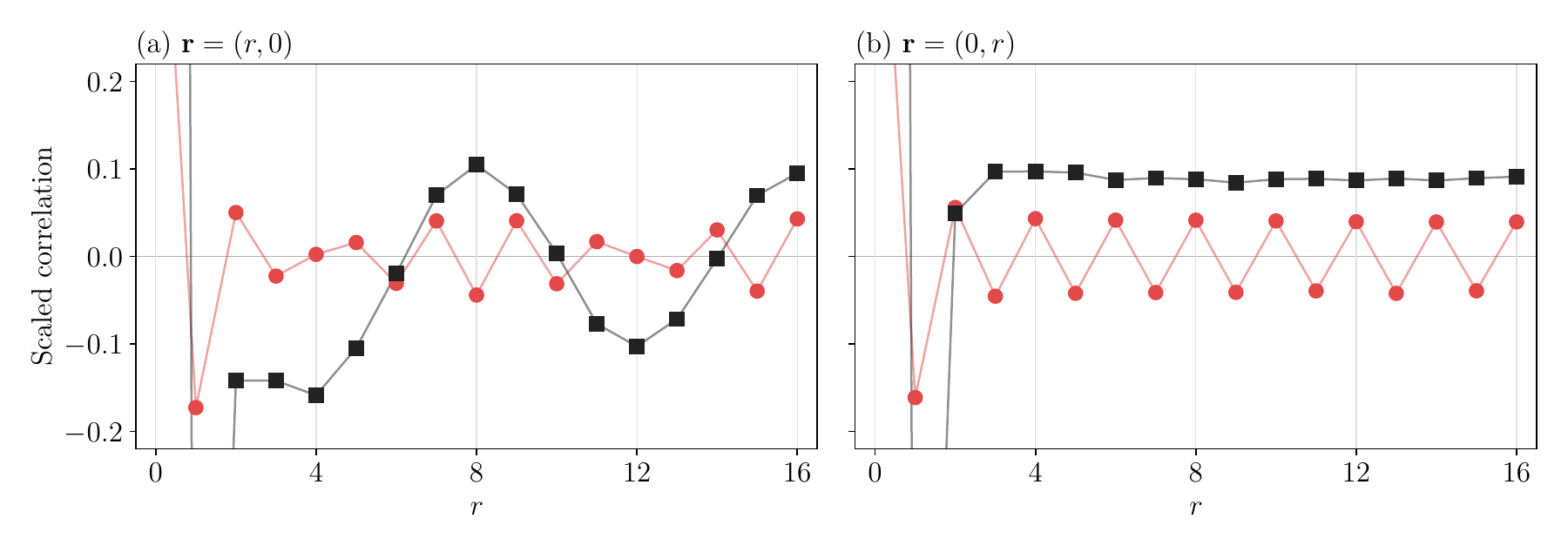}
    \caption{
        Same as Fig.\,\ref{fig:sm_corr_tp0_16_bcs}, for the projected NN-fPEPS on the $32 \times 32$ lattice at $t' = 0$, without $C_p(r)$.
    }
    \label{fig:sm_corr_tp0_32_nn}
\end{figure}

\begin{figure}[tp]
    \centering
    \includegraphics[width=\textwidth]{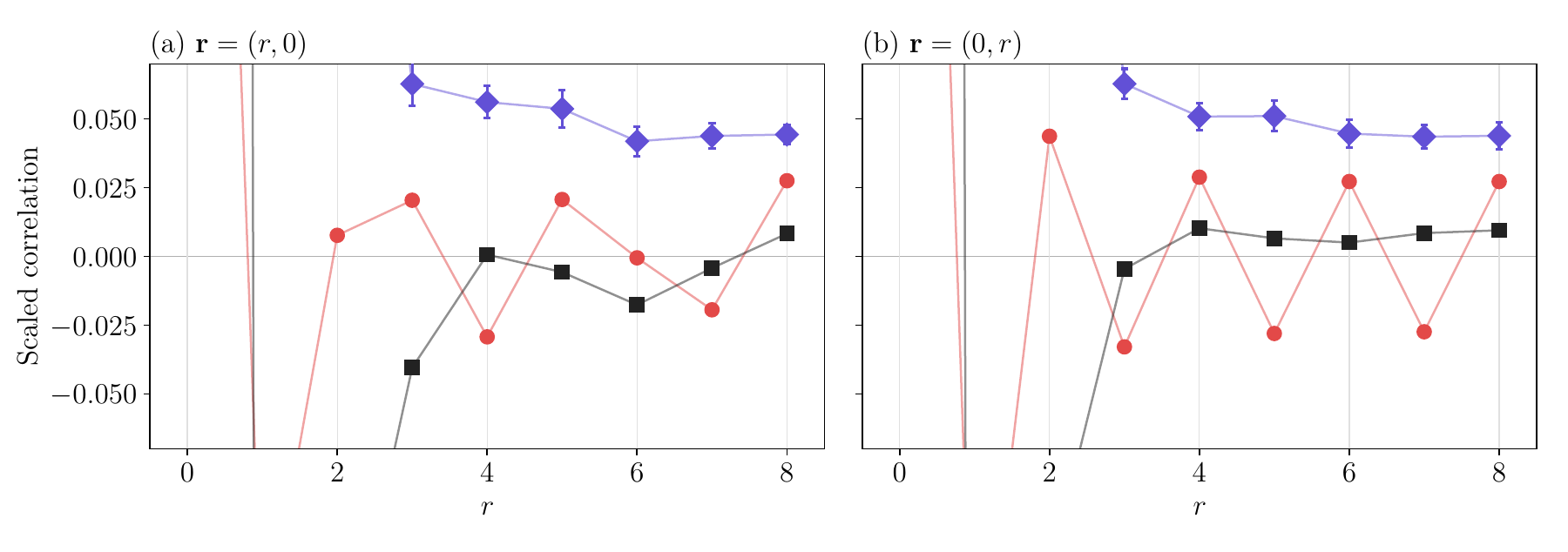}
    \caption{
        Same as Fig.\,\ref{fig:sm_corr_tp0_16_bcs}, for the projected Pfaffian backflow state on the $16 \times 16$ lattice at $t' = -0.2$.
    }
    \label{fig:sm_corr_tpm02_16_pf}
\end{figure}

\begin{figure}[tp]
    \centering
    \includegraphics[width=\textwidth]{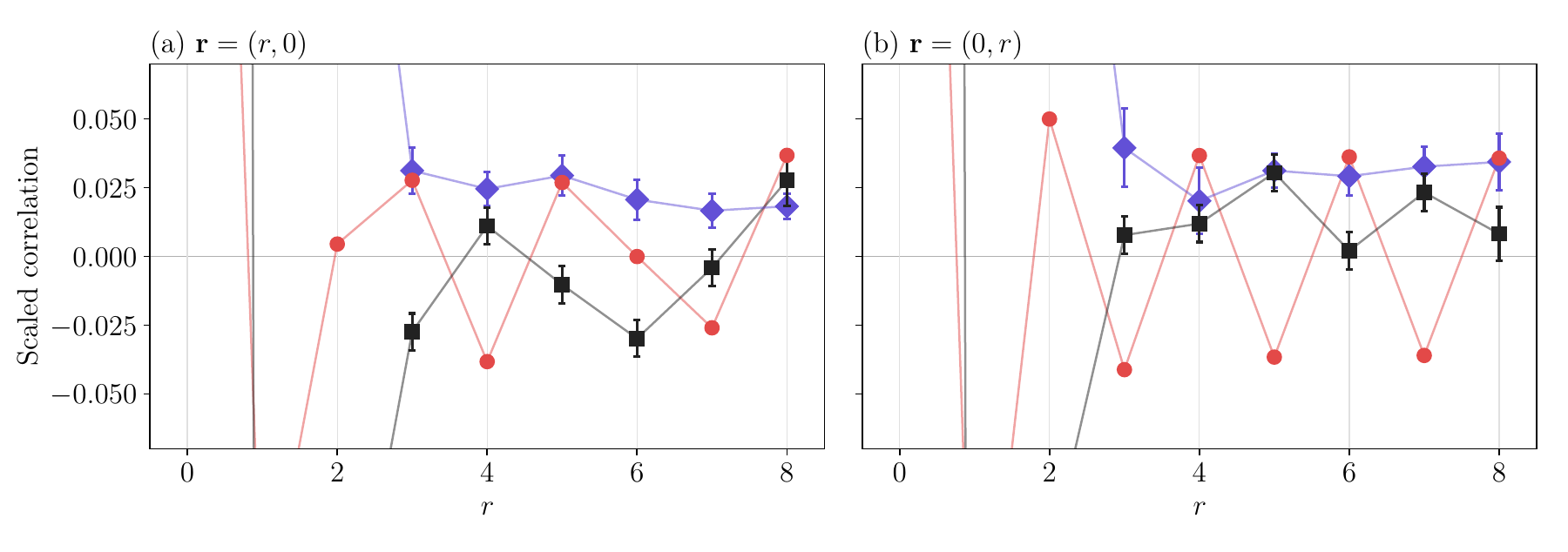}
    \caption{
        Same as Fig.\,\ref{fig:sm_corr_tp0_16_bcs}, for the projected NN-fPEPS on the $16 \times 16$ lattice at $t' = -0.2$.
    }
    \label{fig:sm_corr_tpm02_16_nn}
\end{figure}

\begin{figure}[tp]
    \centering
    \includegraphics[width=\textwidth]{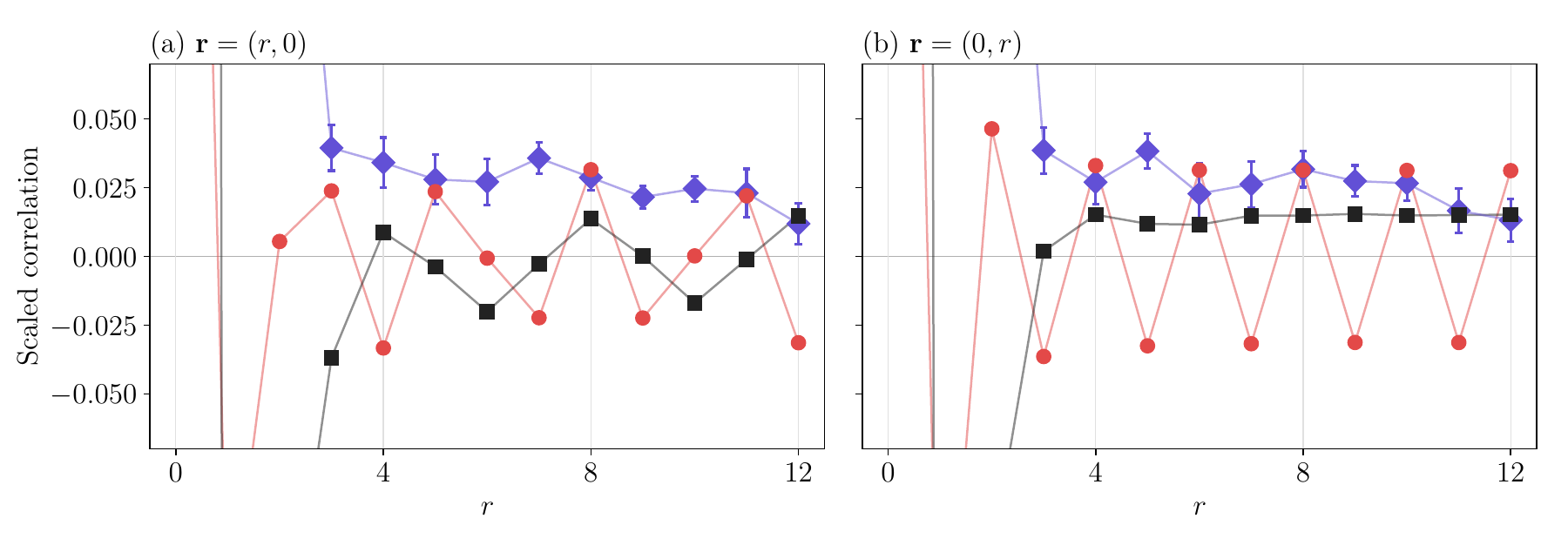}
    \caption{
        Same as Fig.\,\ref{fig:sm_corr_tp0_16_bcs}, for the projected Pfaffian backflow state on the $24 \times 24$ lattice at $t' = -0.2$.
    }
    \label{fig:sm_corr_tpm02_24_pf}
\end{figure}

\begin{figure}[tp]
    \centering
    \includegraphics[width=\textwidth]{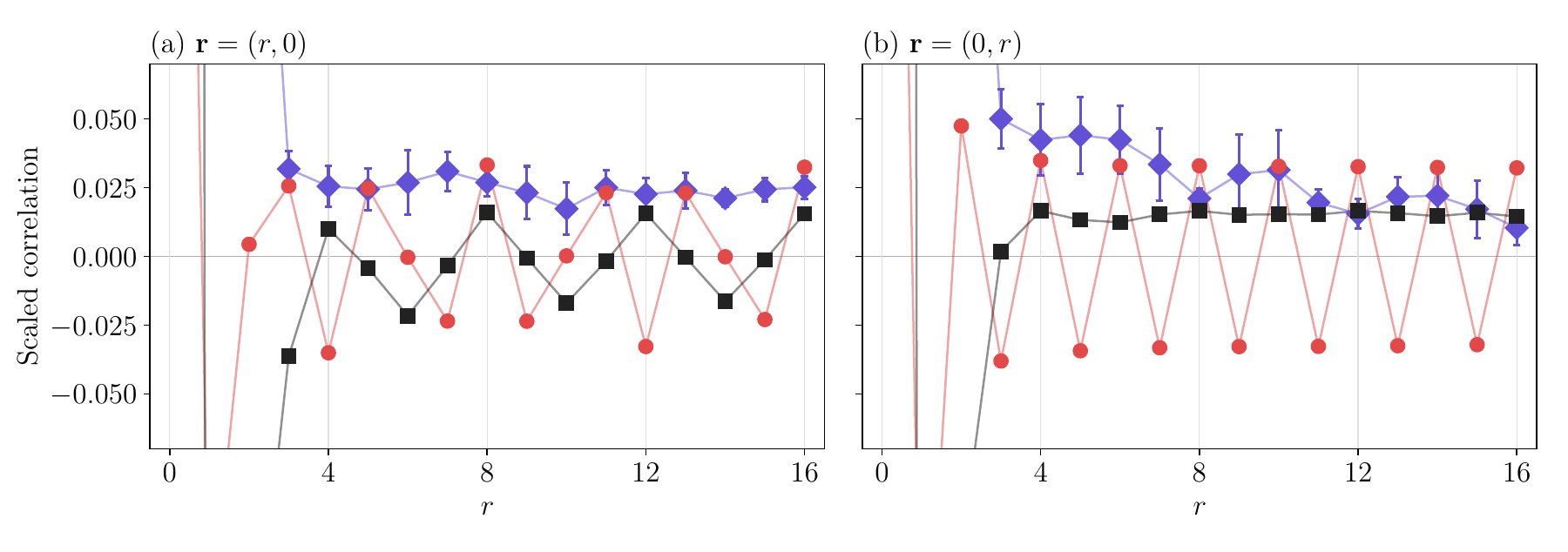}
    \caption{
        Same as Fig.\,\ref{fig:sm_corr_tp0_16_bcs}, for the projected Pfaffian backflow state on the $32 \times 32$ lattice at $t' = -0.2$.
    }
    \label{fig:sm_corr_tpm02_32_pf}
\end{figure}

\begin{figure}[tp]
    \centering
    \includegraphics[width=\textwidth]{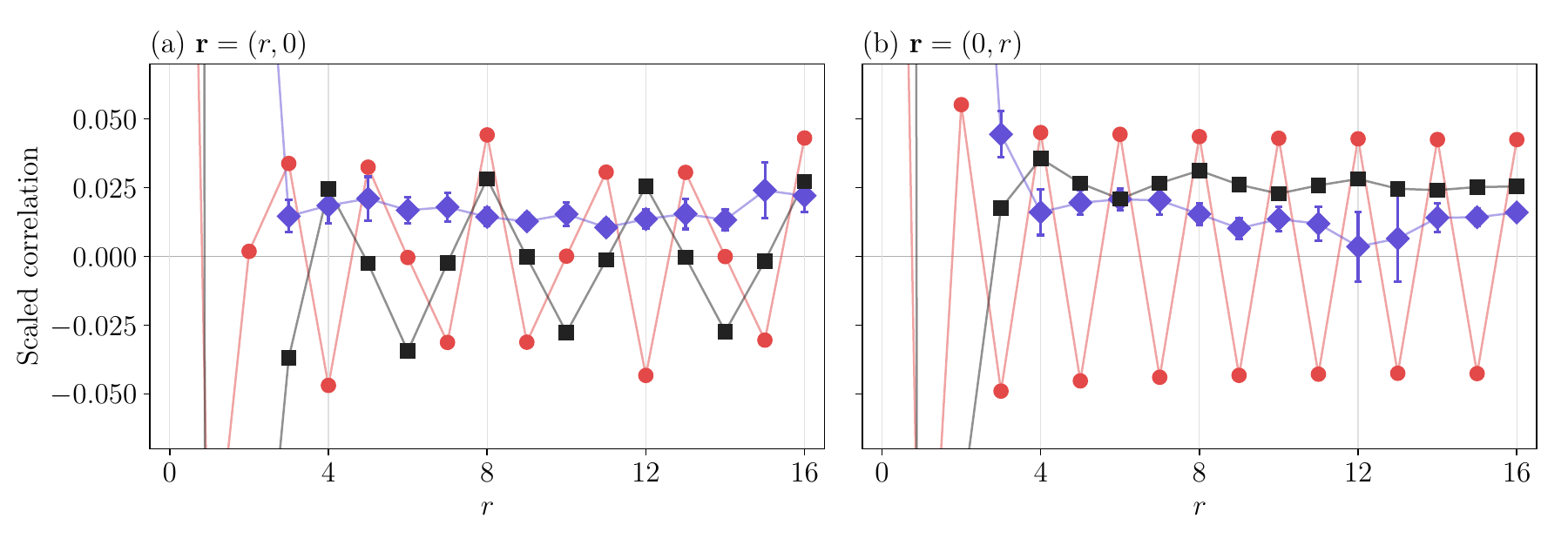}
    \caption{
        Same as Fig.\,\ref{fig:sm_corr_tp0_16_bcs}, for the projected NN-fPEPS on the $32 \times 32$ lattice at $t' = -0.2$.
    }
    \label{fig:sm_corr_tpm02_32_nn}
\end{figure}

Figs.\,\ref{fig:sm_corr_tp0_16_bcs}--\ref{fig:sm_corr_tpm02_32_nn} show the real-space correlations of all our symmetry-projected states on lattices of $16 \times 16$ sites and larger, along the two lattice axes.
Besides the pair-pair correlation $C_p(r)$ of Eq.\,(2) of the main text, they include the spin and charge correlations
\begin{equation}
    C_s(r) = \langle \bm{S}_0 \cdot \bm{S}_r \rangle \,, \qquad
    C_c(r) = \langle (n_0 - \bar{n}) \, (n_r - \bar{n}) \rangle \,,
    \label{eq:sm_corr}
\end{equation}
with $\bar{n} = 7/8$.
To improve visibility of the correlation data, the charge correlation $C_c(r)$ and the pair-pair correlation $C_p(r)$ are multiplied by $50$ and $100$, respectively, to bring them to the same level as the spin correlation $C_s(r)$.

At $t' = 0$, the NN-fPEPS forms a filled stripe with a spin-density wave (SDW) of period $16$ and a charge-density wave (CDW) of period $8$, whose spin and charge correlations persist to the largest distance while its pair-pair correlation decays.
The BCS and Pfaffian backflow states instead have short-ranged spin and charge correlations and a saturated pair-pair correlation.
At $t' = -0.2$, the Pfaffian backflow and the NN-fPEPS both form a half-filled stripe with an SDW of period $8$ and a CDW of period $4$, coexisting with a finite pair-pair correlation that becomes weaker when the system size increases.
The BCS backflow result at $t'=-0.2$ is from Ref.~\cite{Rende_arxiv26_SBP} instead of this work. As the correlations were not provided in the work, we do not show them here.

\clearpage